\documentclass{jfm}

\usepackage{graphicx}
\usepackage{physics}
\usepackage{subcaption}
\usepackage{mathtools}
\usepackage{newtxtext}
\usepackage{newtxmath}
\usepackage{natbib}
\usepackage{hyperref}
\hypersetup{
    colorlinks = true,
    urlcolor   = blue,
    citecolor  = black,
}

\newcommand{\RomanNumeralCaps}[1]
\linenumbers

\title{Poromechanical modelling of responsive hydrogel pumps}

\author{Joseph J. Webber
  \corresp{\email{joe.webber@warwick.ac.uk}}
  \and Thomas D. Montenegro-Johnson}

\affiliation{Mathematics Institute, University of Warwick, Coventry CV4 7AL, UK}

\begin{document}
\maketitle

\begin{abstract}
    Thermo-responsive hydrogels are smart materials that rapidly switch between hydrophilic (swollen) and hydrophobic (shrunken) states when heated past a threshold temperature, resulting in order-of-magnitude changes in gel volume. Modelling the dynamics of this switch is notoriously difficult, and typically involves fitting a large number of microscopic material parameters to experimental data. In this paper, we present and validate an intuitive, macroscopic description of responsive gel dynamics and use it to explore the shrinking, swelling and pumping of responsive hydrogel displacement pumps for microfluidic devices. We finish with a discussion on how such tubular structures may be used to speed up the response times of larger hydrogel smart actuators and unlock new possibilities for dynamic shape change.
\end{abstract}

\begin{keywords}
Authors should not enter keywords on the manuscript, as these must be chosen by the author during the online submission process and will then be added during the typesetting process (see \href{https://www.cambridge.org/core/journals/journal-of-fluid-mechanics/information/list-of-keywords}{Keyword PDF} for the full list).  Other classifications will be added at the same time.
\end{keywords}

\section{Introduction}
\label{sec:introduction}
Hydrogels are soft porous materials comprising a cross-linked, hydrophilic, polymer structure surrounded by adsorbed water molecules that are free to move through the porous scaffold \citep{Doi2009GelDynamics, Bertrand2016DynamicsGel}. Though simple in structure, their elastic and soft nature, coupled with the ability to change volume to an extreme degree by swelling or drying, affords them a number of uses in engineering, medical sciences and agriculture \citep{Zohuriaan-Mehr2010AdvancesMaterials, Guilherme2015SuperabsorbentReview}. In traditional hydrogels, this swelling and drying occurs passively. In responsive hydrogels, the affinity of the polymer scaffold for water changes as a result of external stimuli such as heat, light or chemical concentration \citep{Neumann2023Stimuli-ResponsiveTomorrow}, allowing for controllable swelling-shrinking cycles. Such `smart' materials with tunable shape-changing behaviour have applications in soft robotics \citep{Lee2020HydrogelRobotics}, microfluidics \citep{Dong2007AutonomousHydrogels}, and in models of biological processes \citep{Vernerey2017TheEnvironment}.

Though responsive gels can react to stimuli of various forms, the most ubiquitous are thermo-responsive gels, where the affinity of the polymer chains for water drops rapidly at a critical temperature $T_C$. Above this lower critical solution temperature (LCST), hydrogen bonds holding the water molecules in place around the polymer chains break and the release of water molecules is entropically favoured. Perhaps the most widely-studied thermo-responsive gel is poly(\textit{N}-isopropylacrylamide) (PNIPAM), which has an LCST that can be tuned to be close to room temperature, and finds a number of medical applications owing to its biocompatibility \citep{Das2015StimulusRelease}. The effect of deswelling is significant, with many such gels exhibiting an order-of-magnitude volume change at $T_C$, opening up the possibility of a number of macroscopic use cases for responsive gels \citep{Voudouris2013MicromechanicsLCST}.

Modelling the dynamics of this shape change is difficult, and is thus often restricted to simpler geometries such as spheres \citep{Tomari1995HysteresisGels}. The typical modelling approach seeks the dependence of the Helmholtz free energy of the gel on the ambient temperature, encoded by the Flory $\chi$ parameter, representing the attraction between water molecules and polymer chains. This parameter typically decreases with increasing temperature \citep{Cai2011MechanicsHydrogels}, but its value is usually deduced from fitting to experimental data \citep{Afroze2000PhaseNetworks}. Accurately determining the $\chi$ parameter is a long-standing problem in polymer physics, with experimental approaches often difficult, owing to the number of different physical processes underpinning solvent--polymer and polymer--polymer interactions, with some more recent work using machine learning approaches \citep{Nistane2022EstimationLearning} to seek patterns in the variation of $\chi$ with polymer structure. Difficulties are further compounded by the fact that small changes in $\chi$ can lead to large differences in the physics of hydrogels \citep{Afroze2000PhaseNetworks}.

The Helmholtz free energy is then minimised with respect to deformation, determining the equilibrium swelling state at a fixed temperature. However, describing the transient evolution of the state of the hydrogel as the temperature is varied is significantly more difficult, and requires the separate consideration of chemical potentials, polymer network elasticity and induced interstitial flows through the gel.

In classic large-strain poroelastic models \citep{Bertrand2016DynamicsGel}, the principal stresses (in the directions of the principal stretches) are deduced from the energy. These stresses are then balanced with gradients in chemical potential to describe the poroelastic flow, and thus the gel dynamics. Whilst effective, these models rely on a characterisation of the material in terms of a large number of microscopic parameters, are computationally expensive, and result in a series of coupled partial differential equations for porosity, chemical potential and stresses, which potentially masks some of the key macro-scale physics driving the responsive dynamics and offers limited potential for analytical solutions.

It is also possible to model the behaviour of deformable soft porous media using the theory of linear poroelasticity, characterising the gel by its elastic moduli and describing the flow through the scaffold using Darcy's law \citep{Doi2009GelDynamics}. These models are inherently macroscopic, and offer the benefit of analytic tractability. However, they cannot cope with nonlinearities that arise from large swelling strains, and are therefore unsuitable for modelling super-absorbent gels, where the volumetric changes involved in swelling and drying may be of the order of $10$ to $100$ times \citep{Bertrand2016DynamicsGel}.

In this work, we therefore seek a model based only on macroscopically-measurable material properties that can also incorporate large swelling strains and give faster predictions to describe the transient swelling--deswelling states in response to temperature changes. Such a model would be valuable, as it could provide rapid quantitative design input to experimentalists working on applications of responsive gels such as small microfluidic devices \citep{Harmon2003AHydrogels} and robotic actuators \citep{Lee2020HydrogelRobotics}.
 
A macroscopic continuum-mechanical model for passive gels was recently provided by \citet{Webber2023AGels} and \citet{Webber2023AFormulation}. The model allows for nonlinearities in the isotropic strain, whilst linearising around small deviatoric strains. This assumption is equivalent to the statement that, at any swelling state, the hydrogel material acts as a linear-elastic bulk solid, and it reduces the gel dynamics to a nonlinear advection-diffusion equation for the local polymer (volume) fraction $\phi$.  In this paper, we extend this model to incorporate thermo-responsive effects, by assuming that the osmotic pressure (and potentially other material parameters) can depend also on temperature. This dependency leads to different swelling behaviour as the temperature is varied, and different equilibrium states either side of the LCST.

Our model makes the analysis of complicated responsive actuators more tractable, and provides good qualitative and quantitative predictions of the key physics at play. It is broadly applicable to a range of hydrogel actuators in microfluidic devices, such as valves \citep{Dong2007AutonomousHydrogels}, passive pumps (drawing in water through their swelling behaviour) \citep{Seo2019HydrogelRate}, and the displacement pumps \citep{Richter2009MicropumpsHydrogels} that we will consider herein. 

In this paper, we analyse the contraction of a hollow tube formed of thermo-responsive hydrogel when a heat pulse is applied, and, using the thermo-responsive linear-elastic-nonlinear-swelling model derived in section \ref{sec:lens}, we deduce both the shrunken geometry and the transition from swollen to shrunken states by the flow of water through the hydrogel walls and the hollow lumen of the `pipe'.

Notably, we show that the presence of a fluid-filled pore in the centre of a tube enables much faster responses to changes in temperature than in a pure gel, since the flow that results from deswelling is not restricted by viscous resistance through the pore matrix. Our model also gives expressions for the pumping rate and characteristics of the induced peristaltic fluid flow in response to propagating heat pulses.

Finally, we note that in addition to applications driving fluid flow in microfluidic devices, a number of existing applications depend on the ability to tune response times to external stimuli \citep{Maslen2023AMicroactuators}. In such constructions, anisotropic shape changes result from isotropic deswelling that occurs at different rates -- so-called ``dynamic anisotropy'' -- in response to a heat pulse. This behaviour is key to unlocking non-reciprocal shrinking-swelling dynamics, critical for achieving work in the inertialess fluid regime. The existence of a simplified, analytic, understanding of thermo-responsive gels allows us to tune the thickness of the pipe walls to give a desirable response time, affording us predictions for the construction of responsive hydrogel devices with controllable response rates to external stimuli, irrespective of the intrinsic material response rate. We begin with the derivation of the governing equations that would underpin the responses of such devices.

\section{Thermo-responsive linear-elastic-nonlinear-swelling model}
\label{sec:lens}
\begin{figure}
    \centering
    \begin{subfigure}[t]{0.33\linewidth}
        \centering
        \caption{Reference state}
        \includegraphics[width=0.5\linewidth]{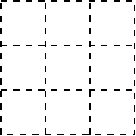}
    \end{subfigure}
    \begin{subfigure}[t]{0.66\linewidth}
        \centering
        \caption{Stress decomposition}
        \includegraphics[width=0.925\linewidth]{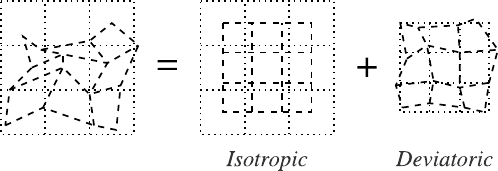}
    \end{subfigure}
    \vspace{0.5em}
    \caption{On the left, the reference state where $\phi \equiv \phi_0$ and the cross-linked polymers are in thermodynamic equilibrium with the surroundings. On the right, a schematic decomposition of any deformation (dashed lines) from this reference state (dotted lines) into an isotropic part due to drying (in this case) and a small deviatoric part.}
    \label{fig:lens_schematic}
\end{figure}
The linear-elastic-nonlinear-swelling (LENS) model introduced in \citet{Webber2023AGels} and \citet{Webber2023AFormulation} is a poromechanical continuum model for the behaviour of large-swelling gels. The model is derived based upon the assumption that isotropic strains, corresponding to the swelling and drying of a gel, may be large, but deviatoric strains must be small. This model achieves both the accurate description of large deformation seen in nonlinear energy-based models and the analytic tractability of linear poroelasticity. Figure \ref{fig:lens_schematic} shows how a general deformation from a reference state can be decomposed into these two parts, and illustrates how we can view isotropic shrinkage or growth as drying or swelling, respectively, changing the local polymer volume fraction $\phi$. In this model, at any given degree of swelling, a hydrogel is characterised using three swelling-dependent material parameters; a generalised osmotic pressure $\Pi(\phi)$ representing the gel's affinity for water, a shear modulus $\mu_s(\phi)$ representing the resistance to elastic deformation and a permeability $k(\phi)$ representing the ease with which water can percolate through the gel scaffold.

In addition to the comparative simplicity of such an approach, these three parameters correspond to clear physical processes, as opposed to microscopic forces on the scale of the polymer chains, or thermodynamic constants that can be difficult to relate to larger-scale swelling or drying phenomena. For example, when applying a force to a gel, the initial, incompressible, response is mediated by the shear modulus $\mu_s$, the final steady state as water is driven in or imbibed is set by the generalised osmotic pressure $\Pi$, and the timescale over which this occurs is set by the permeability $k$. These parameters can be determined for any hydrogel without an understanding of the micro-scale structure or the thermal physics governing the osmotic and elastic behaviour of these materials. 

This is of particular use when considering thermo-responsive gels, where a large number of parameters such as the cross-linker density and interaction parameters must be estimated from reference values or curve fitting \citep{Hirotsu1987Volume-phaseGels}. Perhaps the clearest illustration of the distinction between our macroscopic model and models based on micro-scale physics is our generalised osmotic pressure $\Pi$. This differs from the osmotic pressures in the hydrogel literature by also incorporating isotropic elastic stresses on gel elements, as well as the affinity of polymer chains for water \citep{Webber2024DynamicsHydrogels}. Phenomenologically, the two effects are indistinguishable, and lead to expulsion or imbibition of water, even though their physical basis is vastly different \citep{Peppin2005PressureSuspensions}, and we will henceforth refer to our parameter $\Pi$ simply as the \textit{osmotic pressure} for this reason.

\begin{figure}
    \begin{center}
        \includegraphics[width=.95\linewidth]{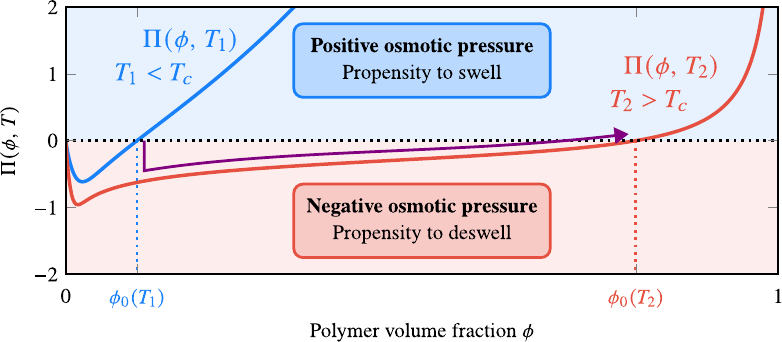}
        \caption{Plots of a representative osmotic pressure function at a temperature $T_1$ below the lower critical solution temperature $T_C$ and $T_2$ above this threshold, showing how the equilibrium polymer fraction increases sharply as the temperature is raised. A sample trajectory as the temperature is raised from $T_1$ to $T_2$ is plotted.}
        \label{fig:phi_0_schema}
    \end{center}
\end{figure}
Thermo-responsive hydrogels respond to changes in temperature through rapidly losing their affinity for water when the lower critical solution temperature (LCST) is exceeded. Macroscopically, this manifests itself as the expulsion of water from the pore spaces and the drying out of the remaining polymer scaffold as a result. This corresponds to a raising of the equilibrium polymer fraction (the polymer fraction attained by a gel placed in water with no external constraints) $\phi_0$ as a result of the change in temperature, and we incorporate this effect into a LENS model by introducing a temperature-dependent equilibrium polymer fraction $\phi_0(T)$. By definition, the osmotic pressure is zero when $\phi=\phi_0$, and so in order to accommodate thermo-responsivity into our model, we must allow the osmotic pressure to depend on $T$, with
\begin{equation}
    \Pi = \Pi(\phi,\,T) \qq{and} \Pi(\phi_0(T),\,T) = 0.
\end{equation}
In figure \ref{fig:phi_0_schema} we illustrate two potential forms of the osmotic pressure below and above the critical temperature $T_C$, and show the mechanism for transition between the equilibrium swelling states $\phi_0(T_1)$ and $\phi_0(T_2)$ as the temperature is raised and a hydrogel deswells.

In experiments, it is observed that the equilibrium polymer fraction rises rapidly as the threshold $T=T_C$ is crossed, with little variation in $\phi_0(T)$ either side of this critical temperature \citep{Butler2022TheHydrogels}. This motivates the choice of a piecewise constant equilibrium polymer fraction
\begin{equation}
    \phi_0(T) = \begin{dcases}\phi_{00} &T \le T_C \\ \phi_{0\infty} & T > T_C\end{dcases},
    \label{eqn:piecewise_phi0}
\end{equation}
where $\phi_{0\infty}$, the `deswollen' equilibrium polymer fraction, is greater than that in the `swollen' state, $\phi_{00}$. The simplest continuous osmotic pressure functions that capture positivity above the equilibrium and negativity below it are defined by
\begin{equation}
    \Pi(\phi,\,T) = \begin{dcases}\Pi_{00}\frac{\phi-\phi_{00}}{\phi_{00}} & T \le T_C \\\Pi_{0\infty}\frac{\phi-\phi_{0\infty}}{\phi_{0\infty}} & T > T_C \end{dcases},
    \label{eqn:piecewise_osmotic_pressure}
\end{equation}
akin to the linearised osmotic pressures used in \citet{Webber2023AGels}, with the parameters $\Pi_{00}$ and $\Pi_{0\infty}$ representing the strength of the osmotic pressures when the polymer fraction is perturbed from its equilibrium value. In the present study, we use the expressions of equations \eqref{eqn:piecewise_phi0} and \eqref{eqn:piecewise_osmotic_pressure} for their analytic simplicity and their ability to capture the macroscopic deswelling behaviour as the LCST is crossed, but in principle any expression for the osmotic pressure can be substituted into the LENS model. Indeed, in appendix \ref{app:energy_based}, we illustrate how LENS parameters can be deduced from a standard model for thermo-responsive gels, employing a neo-Hookean elastic model for the polymer chains and Flory-Huggins theory for the mixing of water and polymer molecules \citep{Cai2011MechanicsHydrogels}.

To model the stresses and strains on a hydrogel element, we measure the displacement $\boldsymbol{\xi}$ from a fixed reference state. In \citet{Webber2023AGels}, this was chosen to be the ``fully swollen'' equilibrium state $\phi\equiv\phi_0$, but we must pick a temperature-independent reference when gels are thermo-responsive. We choose some reference temperature $T_0 < T_C$ where $\phi_0=\phi_{00}$ and consider this the fully-swollen reference state relative to which all displacements are measured. Therefore, the Cauchy strain is equal to
\begin{equation}
    \mathsfbi{e} = \left[1-\left(\frac{\phi}{\phi_{00}}\right)^{1/3}\right]\mathsfbi{I} + \mathsfbi{\epsilon},
\end{equation}
with $\mathsfbi{\epsilon}$ the traceless deviatoric strain, assumed small in LENS modelling. This shows that the divergence of the displacement field satisfies
\begin{equation}
    \bnabla\bcdot\boldsymbol{\xi} = 3\left[1-\left(\frac{\phi}{\phi_{00}}\right)^{1/3}\right].
    \label{eqn:div_xi}
\end{equation}
The stresses on an element of gel are given by the Cauchy stress tensor
\begin{equation}
    \mathsfbi{\sigma} = -\left[p+\Pi(\phi,\,T)\right]\mathsfbi{I} + 2\mu_s(\phi)\mathsfbi{\epsilon},
    \label{eqn:cst}
\end{equation}
where $p$ is the pervadic, or Darcy, pressure (the fluid pressure as would be measured by a transducer separated from the gel by a partially-permeable membrane that only allows water to pass, \citet{Peppin2005PressureSuspensions}). Gradients in pervadic pressure $p$ drive interstitial fluid flows relative to the polymer scaffold, giving a net volume flux $\boldsymbol{u}$ of water via Darcy's law,
\begin{equation}
    \boldsymbol{u} = -\frac{k(\phi)}{\mu_l}\bnabla p,
    \label{eqn:fluid_flux}
\end{equation}
where $k(\phi)$ is the polymer fraction-dependent permeability, which we assume to be equal to a constant $k$ for simplicity. Note that this Darcy velocity is not equal to the water velocity $\boldsymbol{u_w}$ -- instead, it is equal to the flux of water relative to a polymer scaffold that deforms with velocity $\boldsymbol{u_p}$, so $\boldsymbol{u}=(1-\phi)(\boldsymbol{u_w}-\boldsymbol{u_p})$. An expression for the polymer velocity $\boldsymbol{u_p}$ is derived in \citet{Webber2023AFormulation},
\begin{equation}
    \boldsymbol{u_p} = \left(\frac{\phi}{\phi_{00}}\right)^{-1/3}\pdv{\boldsymbol{\xi}}{t},
\end{equation}
representing the reconfiguration of the polymer scaffold as a gel deforms in terms of the displacement field. It can also be shown that the phase-averaged flux $\boldsymbol{q}=\phi\boldsymbol{u_p}+(1-\phi)\boldsymbol{u_w}=\boldsymbol{u}+\boldsymbol{u_p}$ is solenoidal, through conservation of water and polymer.

As shown in the stress tensor of equation \eqref{eqn:cst}, deviatoric elastic strains are related to stresses via the shear modulus $\mu_s(\phi)$, which we henceforth take as a constant $\mu_s$, independent of polymer fraction. This both leads to a more analytically-tractable model, but is also predicted by fully-nonlinear energy-based models, such as those based on neo-Hookean polymer chain elasticity, as outlined in appendix \ref{app:energy_based}.  

Combining the separate expressions for polymer and water conservation with Cauchy's momentum equation, gel dynamics are governed by the polymer fraction evolution equation
\begin{equation}
    \pdv{\phi}{t} + \boldsymbol{q}\bcdot\bnabla\phi = \bnabla\bcdot\left[ D(\phi)\bnabla\phi\right] \qq{with} D(\phi) = \frac{k}{\mu_l}\left[\phi\pdv{\Pi}{\phi} + \frac{4\mu_s}{3}\left(\frac{\phi}{\phi_0}\right)^{1/3}\right].
    \label{eqn:lens_transport}
\end{equation}
The same derivation, presented for example in \citet{Webber2023AGels} and \citet{Webber2024DynamicsHydrogels}, shows that the fluid flux \eqref{eqn:fluid_flux} can be written as $\left[D(\phi)/\phi\right]\bnabla \phi$, such that water flows from areas with low $\phi$ towards drier regions with high $\phi$. Equation \eqref{eqn:lens_transport} can then be coupled with boundary conditions on this flow field and on the stress \eqref{eqn:cst} to solve for the evolution of composition in time. Techniques outlined in \citet{Webber2023AFormulation} allow for the shape of the gel to be deduced from its composition, solving a biharmonic equation for the displacement field $\boldsymbol{\xi}$, but for the simple geometries considered in this paper, equation \eqref{eqn:div_xi} alongside symmetry assumptions will suffice.

\subsection{Comparing thermo-responsive LENS with a fully-nonlinear model}
\label{sec:sphere_drying}
To show that the predictions of LENS modelling compare well with those of commonly-used nonlinear modelling of thermo-responsive hydrogels, we consider the swelling and drying of a poly(\emph{N}-isopropylacrylamide) (PNIPAM) sphere when heated or cooled around its critical temperature $T_C$. This problem has been treated extensively in the literature owing to its geometric simplicity and tractability \citep{Matsuo1988KineticsGels, Tomari1995HysteresisGels,Butler2022TheHydrogels}. Since no constitutive laws for material parameters are specified in LENS, we can deduce functional forms for $\Pi(\phi)$, $\mu_s(\phi)$ and $k(\phi)$ given any model of our choice, including the fully-nonlinear models used by other authors.

Starting from the most common approach of choosing a Gaussian-chain nonlinear elastic model for the polymer scaffold coupled with Flory-Huggins theory for interaction between polymer and water molecules \citep{Cai2011MechanicsHydrogels}, we derive the form of $\Pi(\phi)$ in appendix \ref{app:energy_based},
\begin{equation}
    \Pi(\phi) = \frac{k_B T}{\Omega_f}\left[\Omega^{-1}\left(\phi-\phi^{1/3}\right)-\phi-\operatorname{ln}{(1-\phi)}-\phi^2\chi +\phi^2(1-\phi)\pdv{\chi}{\phi}\right],
\end{equation}
where $k_B$ is the Boltzmann constant, $\Omega_f$ is the volume occupied by a single water molecule, $\Omega$ is the volume of polymer molecules relative to water molecules, and $\chi$ is the Flory interaction parameter, quantifying the affinity of polymer chains for water molecules. This osmotic pressure function permits us to deduce the equilibrium polymer fraction as a function of temperature, with a sharp change at the critical temperature $T_C$. A similar approach gives the shear modulus $\mu_s$, that is found to be independent of polymer fraction. Then, we fit the measured parameters of \citet{Hirotsu1987Volume-phaseGels} to find the osmotic modulus, shear modulus and permeability for such gels in our formalism, as detailed fully in appendix \ref{app:energy_based} and the supplementary material. This parameter set was chosen to avoid the complicated hysteresis behaviour seen in other such fitting parameters (for example, those measured in \citet{Afroze2000PhaseNetworks}), which can be modelled using our approach but we do not discuss here. A more detailed discussion of differences between swelling and drying and spinodal decomposition between different sets of parameters can be found in \citet{Butler2022TheHydrogels}. Further details of the governing equations in both cases, and the precise forms of the non-dimensional times and lengths $t_{BMJ}$ and $r_{BMJ}$ can be found in the supplementary material. 

\begin{figure}
    \begin{center}
        \subcaptionbox{Swelling dynamics}[0.5\textwidth]
        {\includegraphics[width=0.95\linewidth]{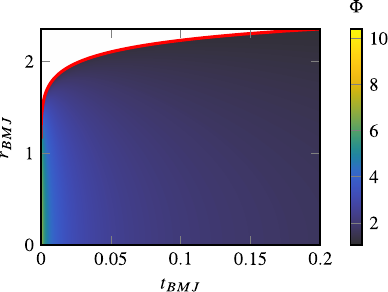}}
        \subcaptionbox{Polymer fraction evolution}[0.44\textwidth]{\vspace{0.35cm}\includegraphics[width=0.93\linewidth]{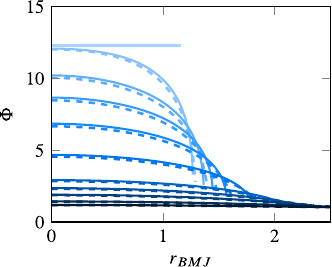}}
        \caption{Plots illustrating the swelling of a hydrogel bead after the temperature is lowered from $308\,\mathrm{K}$ to $304\,\mathrm{K}$. The parameters used are the same as in \citet{Butler2022TheHydrogels}, with the fully nonlinear results plotted for comparison, and $t_{BMJ}$ is the non-dimensional time used by Butler \& Montenegro-Johnson. On the left, the evolving polymer fraction is shown with the growth of the radius in the fully nonlinear model shown as a red curve. On the right, porosity profiles are shown at $t_{BMJ} = 0.0001$, $0.0002$, $0.0005$, $0.001$, $0.0025$, $0.01$, $0.05$, $0.1$, $0.2$, $0.5$ and $1$, with darker blue representing later times. Results from the fully nonlinear model are shown as dashed lines.} 
        \label{fig:swell_bead}
    \end{center}
\end{figure}
\begin{figure}
    \begin{center}
        \subcaptionbox{Shrinking dynamics}[0.5\textwidth]
        {\includegraphics[width=0.95\linewidth]{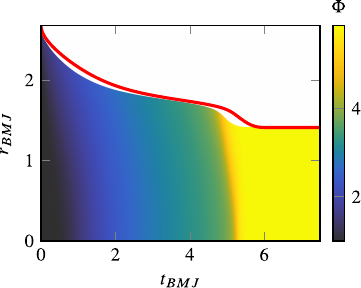}}
        \subcaptionbox{Polymer fraction evolution}[0.46\textwidth]{\vspace{0.325cm}\includegraphics[width=0.93\linewidth]{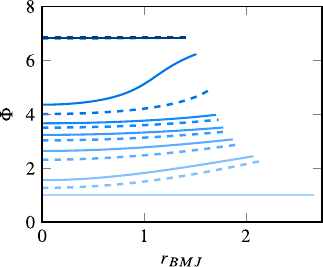}}
        \caption{Plots illustrating the drying of a hydrogel bead after the temperature is raised from $304\,\mathrm{K}$ to $307.6\,\mathrm{K}$, with the same parameters as before and the fully nonlinear solution plotted for comparison. On the left, the evolving porosity is shown with the shrinkage of the radius in the fully nonlinear model shown as a red curve. On the right, porosity profiles are shown at $t_{BMJ} = 0$, $1$, $2$, $3$, $4$, $5$, $6$, $7$ and $8$, with darker blue representing later times. Results from the fully nonlinear model are shown as dashed lines.} 
        \label{fig:dry_bead}
    \end{center}
\end{figure}

We first compute the swelling behaviour of a gel that is initially in equilibrium at $T=308\,\mathrm{K}$ before the temperature is rapidly decreased to $304\,\mathrm{K}$. This leads to swelling from an initial polymer fraction of $\phi_0 \approx 0.64$ to a much lower value $\phi_0 \approx 0.05$. Figure \ref{fig:swell_bead} shows good quantitative and qualitative agreement with the results of \citet{Butler2022TheHydrogels}, with marginally slower growth of the radius but the same diffusive transport of water from surroundings into the bulk of the gel. Repeating the same analysis for smooth drying (where there is no formation of a drying front), we raise the temperature from $304\,\mathrm{K}$ to $307.6\,\mathrm{K}$, with figure \ref{fig:dry_bead} showing the good qualitative, but weaker quantitative, agreement in this case. Our model does, however, capture the rapid initial and later-time drying, with a plateau of slower drying present when $2 \le t_{BMJ} \le 5$.

There is a more significant discrepancy in the predictions of LENS and the fully-nonlinear model in this case due to the significant polymer fraction gradients present close to $t_{BMJ} =5$. In \citet{Butler2022TheHydrogels}, the criteria for gel deswelling with phase separation are deduced, and in this smooth deswelling problem, we pass close to a region of parameter space where phase separation can occur. The presence of a nearby equilibrium solution gives a critical slow-down behaviour akin to that discussed by \citet{Gomez2017CriticalInstabilities}, manifesting itself as the plateau of slow drying at intermediate times.

\subsubsection{Phase separation and negative diffusivities}
Often during the deswelling process a sharp drying front forms, travelling radially inwards through the bead, with the exterior rapidly drying to its final state and the interior remaining relatively swollen until the front reaches the centre. This occurs when trajectories in $(T,\,\phi)$-space pass through the spinodal or coexistence regions. In the spinodal region, spontaneous phase separation can occur, with the formation of regions of dried polymer surrounded by swollen gel or vice versa as the system equilibrates. The coexistence region is a special case of this, where a dried gel and a swollen one can coexist in thermodynamic equilibrium with a simple sharp boundary (such as a drying front) separating the two. In the present study, we consider both of these effects to be forms of spinodal decomposition, with coexistence a weaker `local' form. In either case, there are sharp differences in $\phi$ across very short distances, as seen especially in cases where there is significant hysteretic behaviour in the equilibrium curve (for example in the gel parameters measured by \citet{Afroze2000PhaseNetworks}).

Since large gradients in polymer fraction lead to large deviatoric strains, we expect that our model is unlikely to capture the dynamics of these sharp fronts exactly, since it is dependent on the assumption that these strains remain small. Attempting to replicate this behaviour regardless, through raising the temperature from $304\,\mathrm{K}$ to $308\,\mathrm{K}$, shows that the polymer diffusivity is, in fact, negative for this case in our model. This leads to spinodal decomposition, with $D(\phi)<0$ the criterion for such behaviour to occur.
\begin{figure}
    \centering
    \includegraphics[width=0.95\linewidth]{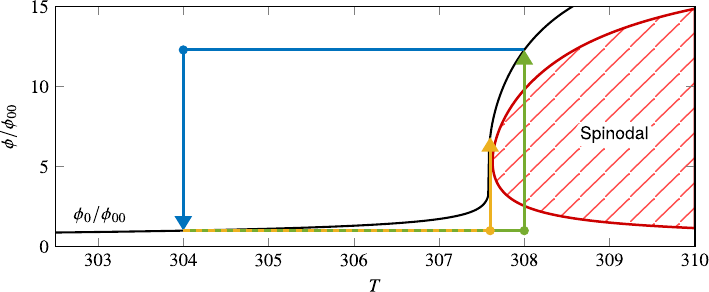}
    \caption{A plot of the region in $(T,\,\phi)$-space where the polymer diffusivity is negative (spinodal region), alongside the equilibrium polymer fraction $\phi_0(T)$, using the gel parameters of \citet{Hirotsu1987Volume-phaseGels} The smooth swelling problem of figure \ref{fig:swell_bead} is plotted in blue, with the temperature lowered and the spinodal region never approached, and the smooth drying of figure \ref{fig:dry_bead} is plotted in yellow. Phase separation occurs, for example, when the temperature is raised to $308\,\mathrm{K}$ and the path to equilibrium passes through the spinodal region, as shown in the example green trajectory.}%
    \label{fig:spinodal}%
\end{figure}%
Figure \ref{fig:spinodal} shows the trajectory of swelling and drying problems in $(T,\,\phi)$-space for one particular choice of parameters, making it clear why swelling (when the temperature is lowered) never leads to negative diffusivities, and why some drying can occur (such as that of figure \ref{fig:dry_bead}) without entering the spinodal region. In the remainder of this paper, we will consider cases of smooth drying where phase separation does not occur: indeed, taking the linear form of $\Pi$ in \eqref{eqn:piecewise_osmotic_pressure} enforces this.

\section{Response times and flow in thermo-responsive tubes}
The transport equation \eqref{eqn:lens_transport} illustrates how the time for a gel to respond to a change in the local temperature is set by the poroelastic timescale $t_{\text{pore}}$ for the gel, found by scaling terms in the equation to be given by
\begin{equation}
    t_{\text{pore}} \sim \frac{\mu_l L^2}{k \Pi_{0\infty}},
    \label{eqn:poroelastic_timescale}
\end{equation}
where $L$ is a lengthscale for the problem. For example, in the plots of figure \ref{fig:swell_bead} where $L \sim a_0$, we see that a swelling sphere only attains its final radius at a time $O(\mu_l a_0^2/k\Pi_{0\infty})$ after the temperature has been changed. In general, these timescales are slow, of the order of many hours for most macroscopic gels of interest \citep{Webber2023AGels}, since the response is rate-limited by the permeability $k$, typically of the order $10^{-15}\,\mathrm{m}^2$ or smaller \citep{Etzold2021TranspirationHydrogels}. 

If the physical situation we are modelling has a fixed size $L$, we seek an approach to lower the poroelastic timescale so that the gel reacts more quickly. Recently, a new class of microfluidic actuators have been designed, reliant on simple geometric designs to convert the isotropic shrinkage of hydrogels above the LCST threshold into more complicated anisotropic morphological changes \citep{Maslen2023AMicroactuators}. Even at the micrometre scale, these devices take a number of seconds to pass through a single actuation cycle, and with deswelling times scaling like $L^2$, centimetre- or millimetre- scale devices harnessing the same physics can be expected to take many hours to achieve the same shape changes. This currently confines such applications to microfluidics, whilst an approach that lowers the response times could find applications in actuators or soft robotics on the macroscopic scale. Recent technical developments have centred on engineered structures with interconnected microchannels that respond much faster to changes in temperature, but detailed modelling of these effects has not been carried out \citep{Spratte2022ThermoresponsiveMicrochannels}.

Concurrently, a number of recent advances in microfluidics have harnessed the ability of hydrogels to pump fluid, either passively through their hydrophilic nature \citep{Dong2007AutonomousHydrogels}, or through the use of responsive hydrogels to drive peristaltic flows \citep{Richter2009MicropumpsHydrogels}. In this latter case, fluid flows many orders of magnitude faster than the percolating flow through the gel matrix can be achieved by squeezing water through microscale voids in the structure. In this section, we consider the simplest such pumping device: a hollow tube of thermo-responsive hydrogel filled with and surrounded by water, and how the tube responds to an increase in temperature above the LCST. This provides a foundation for understanding more complicated physical situations -- for example, understanding the behaviour of such a tube enables the modelling of a single microchannel in microporous gels, allowing for quantitative modelling of response times when such gels are heated.

\subsection{Model problem}
We consider an infinite tube, symmetric around $z=0$, formed from thermo-responsive gel, occupying the region $a_0 < r < a_1$. The lumen of this tube is filled with water and it is surrounded by water. Initially, the gel is in a swollen state with uniform polymer fraction $\phi \equiv \phi_{00}$ and the temperature is constant everywhere, equal to $T_C - \Delta T$, below the critical threshold for deswelling. When the temperature is brought above the critical value, the gel will deswell, leading to a shrinkage of the tube, and the expulsion of water. This water can be expelled radially out of the tube, carried (slowly) through the gel parallel to the axis, or can be transported axially in the lumen of the tube. Though the deswelling response to the temperature change is still governed by the poroelastic timescale, the tube can be manufactured to be sufficiently thin that shrinkage is rapid, and bulk water can be transported much more rapidly through the hollow lumen than would otherwise be the case for a solid cylinder (as in the case investigated by \citet{Webber2023AFormulation}), so that the gel device acts like a small-scale displacement pump, reacting on a much faster timescale than $t_{\text{pore}}$.

\begin{figure}
    \centering
    \includegraphics[width=0.8\linewidth]{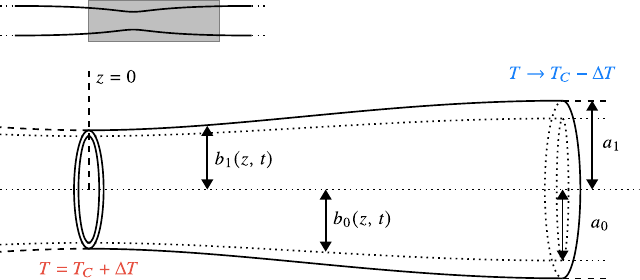}
    \vspace{1em}
    \caption{An illustration of a section of the hydrogel tube in $z\ge 0$, occupying the region $b_0 < r < b_1$ with a hollow lumen inside. The heat pulse that starts at $z=0$ has spread out here leading to a collapse of the tube, which is fully swollen as $z \to \infty$. At temperatures below the LCST, the tube occupies its original position $a_0 < r < a_1$.}
    \label{fig:schematic_tube}
\end{figure}
The deswelling of tubes formed from hydrogels has been studied in the past, specifically in the context of water-filled tubes exposed to the air, losing water through their walls as they dry out \citep{Curatolo2023CircumferentialDehydration}. Qualitatively, many of the same phenomena as seen in our model situation are seen here: water is driven radially from a fluid-filled pore, through the thin walls of the tube, and then out into the surroundings as the gel deswells. However, notably, our gels are surrounded by water and not air, so the tube is unstressed, since we are effectively imposing zero external chemical potential on the outside of the tube by taking $p=0$ here. This implies that we do not expect to see the strong suction effects seen in air drying, where negative inner pressures arise, which have been shown to lead to a circumferential buckling instability \citep{Curatolo2018DrivingCavity}. In our case, we can therefore assume that the shape of the tube will remain axisymmetric for all time.

In order to illustrate the response of the tube to a temperature field that varies in space and time, we impose a temperature $T_C + \Delta T$ at $z=0$ for all times $t\ge 0$. As this heat pulse spreads out $z \to \pm \infty$ in space, there is a collapse of the tube in regions with $T > T_C$, and water is driven out into the surroundings and towards the still-swollen sections of tube. The radius of the inner lumen is described by $r=b_0(z,\,t)$ whilst the outer radius is $r=b_1(z,\,t)$, such that the collapsed tube occupies the region $b_0 < r < b_1$ and $b_i(z,\,0)=a_i$ ($i=0,\,1$). Figure \ref{fig:schematic_tube} illustrates a section of tube some time after the heat pulse has spread out from $z=0$, showing collapse behind the heat pulse front. Symmetry arguments imply that we can restrict our attention to $z \ge 0$ in most cases, with $b_0$, $b_1$, $T$ and $\phi$ all even functions of $z$.

\subsection{Deformation of the tube}
In line with LENS modelling, we assume that all deformation is locally isotropic, and that deswelling leads to a displacement field (relative to the initial state) with axial component $\eta$ and radial component $\xi$ given by
\begin{equation}
    \frac{\xi}{r} \approx \pdv{\xi}{r} \approx \pdv{\eta}{z} \approx 1 - \left(\frac{\phi}{\phi_{00}}\right)^{1/3}.
\end{equation}
Making this assumption requires the polymer fraction field to be independent of $r$ at leading order, an assumption that is reasonable to make in the slender limit of a tube with much larger horizontal lengthscale than diameter. Since $\eta = 0$ at $z=0$, owing to symmetry around this point, the leading-order displacement field is
\begin{equation}
    \xi = \left[1 - \left(\frac{\phi}{\phi_{00}}\right)^{1/3}\right]r \qq{and} \eta = \int_0^z{\left[1 - \left(\frac{\phi}{\phi_{00}}\right)^{1/3}\right]\,\mathrm{d}u}.
    \label{eqn:dispfield}
\end{equation}
Using this expression for $\xi$ allows us to write
\begin{equation}
    \frac{b_0}{a_0} \approx \frac{b_1}{a_1} \approx \left(\frac{\phi}{\phi_{00}}\right)^{-1/3},
\end{equation}
and so the local thickness of the tube is proportional to $\phi^{-1/3}$.

\subsection{Response to changes in temperature}
In order to derive the response of the tube to changes in temperature, it is necessary to solve for the evolution of the gel composition $\phi(r,\,z,\,t)$. Initially, $\phi \equiv \phi_{00}$ everywhere and the geometry of the problem shows that
\begin{equation}
    \left.\pdv{\phi}{z}\right|_{z=0} = 0 \qq{and} \left.\pdv{\phi}{z}\right|_{z\to\pm\infty} = 0,
\end{equation}
arising from the symmetry of the tube and heat pulse around $z=0$ and the assumption that there is no axial flow through the tube itself (proportional to $\pdv*{\phi}{z}$ through equation \eqref{eqn:fluid_flux}) as $z \to \pm \infty$. Indeed, we can simplify the problem further using the symmetry around $z=0$, and instead solve for the composition in $z \ge 0$ alone, reflecting our solution to extend to negative $z$ values.

There has recently been much discussion on the boundary conditions to be applied at the interface of a gel with its surroundings \citep{Xu2022ComparisonInterface,Xu2024TheoryFlow}. Significantly, it has been shown that the nature of the tangential stress and velocity boundary conditions can have a significant effect on the dynamics of flows within and without a hydrogel. In addition, there are potential frictional effects as fluid flows cross the water--gel boundary. When flows are significant, and the external fluid cannot be assumed quiescent, it is important to choose boundary conditions with care, but in the present study the poroelastic timescale is sufficiently long that viscous stresses are negligible \citep{Webber2023AGels} and the dominant flows are radial, with only small tangential components, so the external fluid is treated as a quiescent bath (even though there are flows, for example, along the axis through the lumen).

At the surface $r=b_1(z,\,t)$, we assume that there is no radial stress exerted by the tube on its surroundings (and vice versa), so $\mathsf{\sigma}_{rr} = 0$. Further assuming that large-scale flows in the water bath surrounding the tube are small, we take the fluid pressure to be a constant $p\equiv 0$ outside of the tube. Continuity of pervadic pressure then implies that $p=0$ on $r=b_1$, and this combines with the condition on $\mathsf{\sigma}_{rr}$ to give
\begin{equation}
    \left.\Pi(\phi)\right|_{r=b_1} = 2\mu_s \epsilon_{rr} = 0 \qq{so} \phi(b_1(z),\,z,\,t) = \phi_0,
    \label{eqn:b1bc}
\end{equation}
since the deviatoric strain is zero by our assumption of local isotropy.

On the inside of the tube, we cannot \textit{a priori} make the assumption of uniform zero pervadic pressure since viscous stresses arising from the lumen flows should be balanced by gradients in $p$. As discussed in appendix \ref{app:coupling}, there is an order-$(a_1/L)^2$ correction to the interior pressure field, leading to a mixed boundary condition with a contribution from $\pdv*{\phi_2}{R}$. However, the viscous stresses are much larger on the interior of the gel than on the exterior, owing to the low permeability, and it can be shown, using equation \eqref{eqn:app:full_interior_bc}, that the lumen pressure field has little effect on the gel dynamics. Thence we can use the same boundary condition on the interior of the gel tube as on the exterior, taking
\begin{equation}
    \phi(b_0(z,\,t),\,z,\,t) = \phi_0.
    \label{eqn:b0bc}
\end{equation}

To describe the evolution of polymer fraction in time as the gel expels water, equation \eqref{eqn:lens_transport} becomes
\begin{align}
    \pdv{\phi}{t} + \boldsymbol{q}\bcdot\bnabla\phi = \frac{1}{r}\pdv{r}&\left[rD(\phi, T)\pdv{\phi}{r}\right] + \pdv{z}\left[D(\phi, T)\pdv{\phi}{z}\right] \quad \text{with} \notag \\ &D(\phi, T) = \frac{k}{\mu_l}\left[\frac{\Pi_0(T)\phi}{\phi_0(T)} + \frac{4\mu_s}{3}\left(\frac{\phi}{\phi_{00}}\right)^{1/3}\right].
    \label{eqn:tube_evolution}
\end{align}

There is no intrinsic axial lengthscale arising from the geometry of this problem, since the tube is infinite in length, but we introduce a lengthscale $L$ representing the characteristic distance over which temperature variations occur. In order to simplify the analysis, we make a slenderness assumption that the characteristic axial lengthscale $L$ is much greater than the characteristic radial lengthscale $a_1$. Define $\varepsilon = a_1/L$, and assume that the polymer fraction field only has leading-order axial variation, with radial differences in polymer fraction being of the order $\delta \ll 1$ (arising from our assumption of local isotropy),
\begin{equation}
    \phi = \phi_1(z,\,t) + \delta \phi_2(r,\,z,\,t),
\end{equation}
where we have made the arbitrary choice that $\phi_2$ is zero on the midline $r=\left[b_0(z,\,t)+b_1(z,\,t)\right]/2$, with $\phi_1$ being the polymer fraction on the middle of the tube. Substituting this form of $\phi$ into the evolution equation \eqref{eqn:tube_evolution} and separating variables, we deduce that $\delta=\varepsilon^2$. Therefore, we need only make a relatively weak slenderness assumption, since only $\varepsilon^2$ need be small.

The material flux $\boldsymbol{q}=q_r\boldsymbol{\hat{r}} + q_z\boldsymbol{\hat{z}}$ is solenoidal and thus $q_r / a_1 \sim q_z / L$, so $q_r \sim \varepsilon q_z$. Therefore,
\begin{equation}
    q_r \pdv{\phi}{r} = \varepsilon^2 q_r \pdv{\phi_2}{r} \sim \frac{\varepsilon^3}{a_1} q_z \qq{and} q_z \pdv{\phi}{z} \sim \frac{\varepsilon}{a_1} q_z,
\end{equation}
allowing us to neglect radial advection using this slenderness assumption. Introducing the non-dimensional variables $R=r/a_1$ and $Z=z/L$, the same non-dimensional scalings can be made to the radii $b_0$ and $b_1$, so
\begin{equation}
    B_0 = \frac{b_0}{a_1} = \ell \left(\frac{\phi_1}{\phi_{00}}\right)^{-1/3} \qq{and} B_1 = \frac{b_1}{a_1} = \left(\frac{\phi_1}{\phi_{00}}\right)^{-1/3},
\end{equation}
where $\ell = a_0/a_1 < 1$. The leading-order balance of equation \eqref{eqn:tube_evolution} is thus
\begin{equation}
    L^2 \pdv{\phi_1}{t} + L q_z \pdv{\phi_1}{Z} = \frac{1}{R}\left[R D(\phi_1,\,T)\pdv{\phi_2}{R}\right] + \pdv{Z}\left[D(\phi_1,\,T)\pdv{\phi_1}{Z}\right].
    \label{eqn:evolution_scaled}
\end{equation}
We now separate variables for $\phi_2$, since the only term that depends on $R$ is the first diffusive term on the right-hand side. Hence,
\begin{equation}
    \phi_2(R,\,Z,\,t) = \frac{f(Z,\,T,\,t)}{4 D(\phi_1,\,T)}R^2 + g(Z,\,T,\,t) \operatorname{ln}{R} + h(Z,\,T,\,t).
\end{equation}
By definition, $\phi_2 = 0$ on the midline of the tube $R=(B_0+B_1)/2$ and $\phi_2 = [\phi_0(T)-\phi_1]/\varepsilon^2$ on $R=B_0$ and $R=B_1$ from boundary conditions \eqref{eqn:b1bc} and \eqref{eqn:b0bc}, so $\phi_2$ is symmetric around the midline, with
\begin{gather}
    \phi_2(R,\,Z,\,t) = \frac{4[\phi_0(T)-\phi_1]}{\varepsilon^2(1-\ell)^2}\left(\frac{\phi_1}{\phi_{00}}\right)^{2/3}\left[R - \frac{1+\ell}{2}\left(\frac{\phi_1}{\phi_{00}}\right)^{-1/3}\right]^2 \quad \text{with} \notag \\ f = \frac{16D(\phi_1,\,T)[\phi_0(T)-\phi_1]}{\varepsilon^2(1-\ell)^2}\left(\frac{\phi_1}{\phi_{00}}\right)^{2/3}.
    \label{eqn:phi2}
\end{gather}

This allows us to deduce the radial structure of the polymer fraction given the value of $\phi_1$, the polymer fraction on the inside of the tube. This is found by solving the evolution equation \eqref{eqn:evolution_scaled}, which is now fully-determined up to the total axial flux. Using the approach outlined in \citet{Webber2023AFormulation},
\begin{align}
    L q_z &= \frac{D(\phi_1,\,T)}{\phi_1}\pdv{\phi_1}{Z} + L \left(\frac{\phi_1}{\phi_{00}}\right)^{-1/3}\pdv{\eta}{t} \notag \\
    &= \frac{D(\phi_1,\,T)}{\phi_1}\pdv{\phi_1}{Z} - \frac{L^2}{3}\left(\frac{\phi_1}{\phi_{00}}\right)^{-1/3}\int_0^Z{\left(\frac{\phi_1}{\phi_{00}}\right)^{-2/3}\pdv{\phi_1}{t}\,\mathrm{d}u}.
\end{align}
This, alongside the form of $\phi_2$ from equation \eqref{eqn:phi2}, can then be substituted into equation \eqref{eqn:evolution_scaled}, which we non-dimensionalise by introducing the variables
\begin{equation}
    \tau = \frac{k\Pi_{00}t}{\mu_l L^2},\; \Phi_{0,\,1,\,2,\,\infty} = \frac{\phi_{0,\,1,\,2,\,0\infty}}{\phi_{00}},\;\mathcal{M} = \frac{\mu_s}{\Pi_{00}},\;\mathcal{D}(\Phi_1,\,T) = \frac{\mu_l}{k \Pi_{00}}D(\phi_1,\,T).
    \label{eqn:nondimensionalisation}
\end{equation}
Then,
\begin{gather}
    \pdv{\Phi_1}{\tau} + \frac{\mathcal{D}}{\Phi_1}\left(\pdv{\Phi_1}{Z}\right)^2 - \frac{\Phi_1^{-1/3}}{3}\pdv{\Phi_1}{Z}\int_0^Z{\Phi_1^{-2/3}\pdv{\Phi_1}{\tau}\,\mathrm{d}u}  = \mathcal{F} + \pdv{Z}\left[\mathcal{D}\pdv{\Phi_1}{Z}\right] \quad \text{with} \notag \\
    \mathcal{D} = \frac{\Pi_0(T)}{\Pi_{00}}\frac{\Phi_1}{\Phi_0(T)} + \frac{4\mathcal{M}}{3}\Phi_1^{1/3} \qq{and} \mathcal{F} = \frac{16 \Phi_1^{2/3}\mathcal{D}[\Phi_0(T)-\Phi_1]}{\varepsilon^2(1-\ell)^2}.
    \label{eqn:tube_model}
\end{gather}
This is to be solved subject to the initial condition $\Phi_1 \equiv 1$ and subject to boundary conditions $\pdv*{\Phi_1}{Z}=0$ at both $Z=0$ and as $Z \to \infty$. In the present study, a finite-difference scheme akin to that summarised in the supplementary material of \citet{Webber2023AFormulation} is used to solve equation \eqref{eqn:tube_model}. From this solution, equation \eqref{eqn:phi2} gives the radial polymer fraction structure, and the shape of the tube is given by
\begin{equation}
    \ell \Phi_1^{-1/3} \le R \le \Phi_1^{-1/3},
\end{equation}
at leading order in the small parameter $\varepsilon^2$. The form of the function $\mathcal{F}$ implies that, for our model to be consistent, $\Phi_1$ must everywhere be close to the piecewise-constant equilibrium polymer fraction $\Phi_0$, or else our scaling arguments for the terms in the advection-diffusion equation will be invalid. We can check this assumption after calculating the solution to verify the validity of our modelling.

\subsection{Response to uniform temperature change}
\label{sec:uniform}
\begin{figure}
    \centering
    \subcaptionbox{Deswelling timescales when $\ell = 0.5$\label{fig:1d_solutions_contour}}[0.49\textwidth]
    {\includegraphics[width=0.95\linewidth]{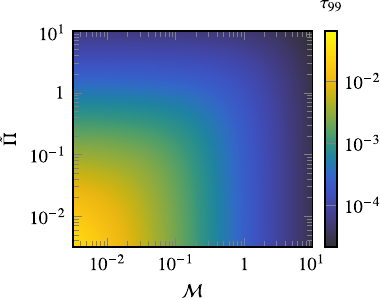}}
    \subcaptionbox{Polymer fraction with $\mathcal{M}=\tilde{\Pi}=1$\label{fig:1d_solutions_trajectories}}[0.49\textwidth]{\vspace{0.38cm}\includegraphics[width=0.82\linewidth]{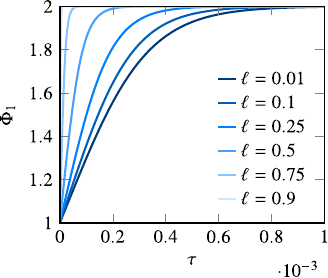}}
    \caption{Plots of the one-dimensional deswelling of a tube when the temperature is uniformly changed when $\Phi_\infty=2$ and $\varepsilon = 0.1$. This shows the variation of the deswelling timescale $\tau_{99}$ (the time taken for $\Phi_1 \ge 1.99$) and the approach to steady state for a number of tube thicknesses.}
    \label{fig:1d_solutions}
\end{figure}
Before studying the response of a hollow tube to a propagating heat pulse, we first consider the case where the temperature is everywhere brought up from below the LCST to $T_C + \Delta T$ at $t=0$. The response of the tube is axially-uniform, evolving following a simplified form of equation \eqref{eqn:tube_model},
\begin{equation}
    \pdv{\Phi_1}{\tau} = \frac{16(\Phi_\infty-\Phi_1)}{\varepsilon^2(1-\ell)^2}\left(\tilde{\Pi}\frac{\Phi_1^{5/3}}{\Phi_\infty} + \frac{4\mathcal{M}}{3}\Phi_1\right),
\end{equation}
where $\tilde{\Pi} = \Pi_{0\infty}/\Pi_{00}$. We can use this equation to understand how the material parameters $\Phi_\infty$, $\mathcal{M}$ and $\tilde{\Pi}$ affect the response time to a change in temperature without the added complication of spatial variations. We know that the polymer fraction on the interior of the tube wall will approach $\Phi_\infty$ as time goes on, with the outside polymer fraction instantaneously reaching this value, but the rate at which this steady state is approached may vary. To measure the rate of deswelling, define the deswelling timescale $\tau_{99}$ as the time taken for
\begin{equation}
    \Phi_1 \ge \Phi^*-\frac{\Phi^*-1}{100}.
\end{equation}

Straightforwardly, it is clear that deswelling is more rapid when there is a greater contrast between $\phi_{00}$ and $\phi_{0\infty}$, since the bracketed term $\Phi_\infty-\Phi_1$ is greater in magnitude. Thus, gels with more dramatic deswelling will approach their steady states faster. Figure \ref{fig:1d_solutions_contour} shows how the time taken to reach $\Phi_\infty$ depends on the stiffness of the gel (encoded by $\mathcal{M}$) and the relative strength of the osmotic pressure at higher temperatures (encoded by $\tilde{\Pi}$). Stiffer gels resist the formation of deviatoric strains, which arise from differences in polymer fraction, so the interior must deswell to catch up with the outside of the tube, leading to a much faster deswelling process as $\mathcal{M}$ increases. Similarly, larger values of $\tilde{\Pi}$ lead to more rapid interstitial flows driven by pervadic pressure gradients, and so the time to deswell decreases as $\tilde{\Pi}$ increases.

Figure \ref{fig:1d_solutions_trajectories} illustrates the approach of the polymer fraction on the interior of the tube wall, $\Phi_1$, to the equilibrium value $\Phi_\infty$, showing how the approach is more rapid for thinner tubes where there is a shorter distance for water to diffuse out.

\subsection{Heat transfer in the system}
If, instead of a uniform temperature field, we impose a fixed temperature $T=T_C + \Delta T$ at $Z=0$, we expect this heat pulse to spread out in the axial direction, symmetrically around the origin, with a deswelling front behind which $T > T_C$, and in front of which $T < T_C$. Modelling the transport of heat through the water, gel scaffold, and within the pore space water, is a potentially complicated task, and a number of different transport processes must be accounted for, as well as the energetic contributions of swelling, deswelling and deformation \citep{Kaviany1995PrinciplesMedia}. In the present study, we will consider the simplest possible case, acknowledging that more complicated phenomena such as dispersion will also contribute to heat transfer, but can be reasonably neglected on the assumption that flows through the gel are sufficiently slow.

There is much discussion in the literature of thermoelasticity with specific applications to hydrogels \citep{Cai2011MechanicsHydrogels,Drozdov2014SwellingHydrogels, Brunner2024NumericalHydrogels}, and here we take a necessarily simpler model, justifying why deformation and swelling do not contribute to temperature evolution at leading order. In appendix \ref{app:thermoelasticity}, we derive a temperature evolution equation for a gel in the LENS formalism in the absence of external heat supply ($R=0$),
\begin{equation}
    \pdv{T}{t} + \boldsymbol{q}\bcdot\bnabla T = \kappa \nabla^2 T + \frac{k(\phi)}{\rho c \mu_l}\left|\bnabla p\right|^2 + \frac{1}{\phi}\left(\frac{\Pi(\phi)}{\rho c} + T\right)\left(\pdv{\phi}{t} + \boldsymbol{q}\bcdot\bnabla\phi\right),
    \label{eqn:lens_heat_equation}
\end{equation}
where $c$ is the specific heat capacity, $\rho$ is the gel density and $\kappa$ is a spatially-averaged thermal diffusivity. In the water surrounding the hydrogel, heat transfer is described by the advection-diffusion equation
\begin{equation}
    \pdv{T}{t} + \boldsymbol{u}\bcdot\bnabla T = \bnabla \bcdot \left(\kappa_w \bnabla T\right) = \kappa_w \nabla^2 T,
\end{equation}
where $\kappa_w$ is the thermal diffusivity of water, assumed to be spatially-uniform. In our model, we assume that $\kappa\approx \kappa_w$. Certainly this is true in regions where $\phi \ll 1$ and the gel remains swollen, since the contributions of diffusivity in the solid are limited here, a statement supported by experiment and molecular dynamics simulation \citep{Xu2018ThermalNanoscale}. In deswollen regions where the polymer fraction is larger, $\kappa$ is likely of the same order of magnitude as $\kappa_w$, since the thermal diffusivity of polymer chains is of the same magnitude as the thermal diffusivity of water \citep{Freeman1987ThermalChain}. However, such regions are a small fraction of the total spatial domain, and the fact that the collapsed tube is thin here means that we neglect any variation from $\kappa_w$ in this region.

There are two potential timescales in the heat transfer problem in the gel -- the poroelastic timescale $t_{\text{pore}}$ of equation \eqref{eqn:poroelastic_timescale}, and the thermal timescale $t_{\text{therm}} = L^2/\kappa$. Their ratio is
\begin{equation}
    \frac{t_{\text{pore}}}{t_{\text{therm}}} = \frac{\kappa}{D} = Le,
\end{equation}
the Lewis number, representing the ratio of thermal to compositional diffusivities. In the case $Le \gg 1$, equation \eqref{eqn:lens_heat_equation} simply reduces to the diffusion equation, since all but the first term on the right-hand side depend on the (slow) reconfiguration of the gel scaffold.

We know that flow and deformation of the gel is mediated by the low permeability of the polymer scaffold, with $k \sim 10^{-15}\,\mathrm{m}^2$, and therefore expect that the transfer of heat by conduction will occur much faster than changes in shape to the gel. The compositional diffusivity $k\Pi_{00}/\mu_l$ typically scales like $10^{-8}\,\mathrm{m}^2\mathrm{s}^{-1}$, whilst $\kappa \sim 10^{-7}\,\mathrm{m}^2 \mathrm{s}^{-1}$, so $Le \sim 10$. In the present study, we restrict our attention to the large-$Le$ limit for simplicity, where heat transfer in both the gel and water can be modelled by
\begin{equation}
    \pdv{\theta}{\tau} = Le \pdv[2]{\theta}{Z} \qq{with} \begin{dcases}
        \theta(0,\,\tau) = 1,\\ \theta \to -1 &\text{as $Z\to\infty$}
    \end{dcases}
    \label{eqn:temp_field}
\end{equation}
where $\theta = (T-T_C)/\Delta T$. There are reasonable physical situations where these assumptions do not apply, but we do not consider them here -- modelling such cases would require a careful consideration of heat transfer by advection, dispersion and diffusion, as well as incorporating the effect of fast fluid flows into boundary conditions at the gel surface \citep{Xu2022ComparisonInterface}. Equation \eqref{eqn:temp_field} has a solution in terms of the error function, with
\begin{equation}
    \theta = 2 \operatorname{erfc}\left(\frac{Z}{2\sqrt{Le\, \tau}}\right) - 1,
    \label{eqn:temp_field_sol}
\end{equation}
where $\operatorname{erfc}$ is the complementary error function \citep{Abramowitz1970HandbookSeries}. In order to understand the response of the gel to the diffusive heat pulse, we first seek the position of the deswelling front $Z=Z_C$ where $\theta=0$. This is found using equation \eqref{eqn:temp_field_sol}, with
\begin{equation}
    \operatorname{erfc}{\left(\frac{Z_C}{2\sqrt{Le\, \tau}}\right)} = \frac{1}{2} \qq{so} Z_C = 2 \operatorname{erfc^{-1}}{\left(\frac{1}{2}\right)}\sqrt{Le\, \tau} \approx 0.9539 \sqrt{Le\,\tau}.
\end{equation}

\subsection{Response to pulses of heat}
\label{sec:heat_pulse}
\begin{table}
    \begin{center}
        \def~{\hphantom{0}}
        \begin{tabular}{ll}
            Parameter  & Value \\[3pt]
            Deswollen scaled polymer fraction $\Phi_\infty$ & $2$ \\
            Ratio of osmotic pressure scales $\tilde{\Pi}$ & $1$ \\
            Aspect ratio $\varepsilon = a_1/L$ & $0.1$ \\
            Shear parameter $\mathcal{M}$ & $1$ \\
            Lewis number $Le$ & $10$
        \end{tabular}
        \caption{Parameter values used in the modelling of drying tubes from section \ref{sec:heat_pulse} onwards, with the effect of changing $\Phi_\infty$, $\tilde{\Pi}$ and $\mathcal{M}$ discussed in section \ref{sec:uniform}.}
        \label{tab:params_tube}
    \end{center}
\end{table}
\begin{figure}
    \centering
    \includegraphics[width=0.95\linewidth]{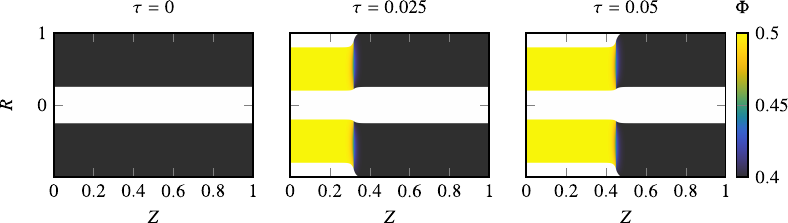}
    \caption{Plots of the evolution of a hollow thermo-responsive hydrogel tube with parameters from table \ref{tab:params_tube} and $\ell = 0.25$. The heat pulse diffuses from left to right, with the gel shrinking behind it.}
    \label{fig:profiles}
\end{figure}
Using the model summarised in equation \eqref{eqn:tube_model}, we can compute the mechanisms by which a thermo-responsive gel tube will collapse in response to a temporally-- and spatially-varying temperature field \eqref{eqn:temp_field_sol}. Key to the behaviour here is the fact that heat diffuses on a faster timescale than the water can diffuse through the polymer, leading to a smooth front centred on the pulse front $Z_C(\tau)$. From this point onwards, we will use the parameters in table \ref{tab:params_tube} in all modelling, having discussed the effect of varying $\mathcal{M}$ and $\tilde{\Pi}$ on the one-dimensional deswelling in previous sections. Figure \ref{fig:profiles} shows the thickness of a tube at different times as heat diffuses and the gel shrinks. Notice that the shrinkage, though rapid, is not instantaneous in time, since the slow diffusion of water out of the walls of the tube sets a delayed response.

\begin{figure}
    \centering
    \subcaptionbox{Relaxation to the equilibrium polymer fraction around $Z_C$\label{fig:relax_profiles}}[0.49\textwidth]
    {\includegraphics[width=0.95\linewidth]{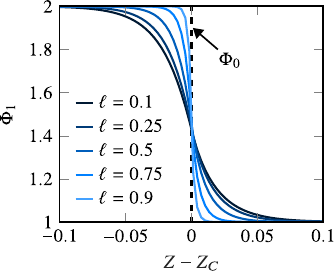}}
    \subcaptionbox{Fitted values of $A(\ell)$ showing that $A \sim (1-\ell)^{-1}$\label{fig:relax_fit}}[0.49\textwidth]{\vspace{0.1cm}\includegraphics[width=0.93\linewidth]{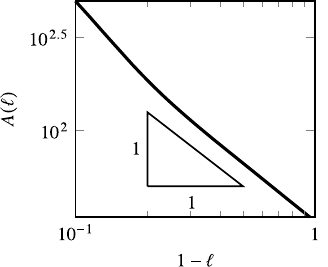}}
    \caption{Plots of the interior polymer fraction $\Phi_1$ at $\tau = 10^{-2}$ with the same parameters as in figure \ref{fig:profiles}, showing how the relaxation to the steady state $\Phi=\Phi_0(T)$ around the drying front $Z=Z_C(\tau)$ is much faster for thinner tubes $\ell \to 1$. These profiles can be approximated by a $\tanh$ function, as in equation \eqref{eqn:tanh_fit}, with fitting parameter $A(\ell)$ shown in the logarithmic plot on the right.}
    \label{fig:relax}
\end{figure}

In order for the gel to deswell, water must flow from the walls of the tube into the surrounding water, the lumen at the centre of the tube, or through the gel itself parallel to the axis. Clearly, if the walls of the tube are thinner, driving water from the hydrogel is more rapid, since the water has less of a distance to diffuse outwards, and we expect a more rapid response to changes in temperature for larger values of $\ell$. The more rapid approach to steady state is shown in figure \ref{fig:relax_profiles}, where the sharper equilibrium profile is approached more closely around the drying front $Z_C(\tau)$ for thinner tube walls. Assuming that the radial fluxes are locally dominant, equation \eqref{eqn:tube_model} reduces to the one-dimensional case of section \ref{sec:uniform},
\begin{equation}
    \pdv{\Phi_1}{\tau} \approx \frac{16\Phi_1^{2/3}\mathcal{D}}{\varepsilon^2(1-\ell)^2} \times \begin{dcases}\Phi_\infty-\Phi_1 \; &Z<Z_C \\ 1-\Phi_1 \; &Z>Z_C\end{dcases},
\end{equation}
away from the front at $Z=Z_C$ (where $\pdv*{\Phi_1}{Z}$ will be significant). Then, timescales decrease like $(1-\ell)^2$ when $\ell$ is increased. In the opposite limit as $\ell \to 0$, adjustment happens on the unmodified poroelastic timescale.

From figure \ref{fig:profiles}, it is clear that the structure of the solution around $Z=Z_C(\tau)$ appears to propagate like a travelling wave centred on the deswelling front, since the contribution of axial flows through the gel is limited compared to that of radial flows. Therefore, we can consider the quasi-one-dimensional problem in the new coordinate $Z-Z_C$. The plots in figure \ref{fig:relax} suggest that polymer fraction can locally be approximated by a smooth step around $Z=Z_C(\tau)$, with the steepness a function of thickness $\ell$. We thus propose that
\begin{equation}
    \Phi_1 \approx \Phi_\infty - \frac{\Phi_\infty-1}{2}\left\lbrace 1 + \tanh{\left[A(\ell)\left(Z-Z_C\right)\right]}\right\rbrace,
    \label{eqn:tanh_fit}
\end{equation}
for some scaling factor $A$, a function of $\ell$, representing the sharpness of the drying front. Figure \ref{fig:relax} shows that $A(\ell) \sim (1-\ell)^{-1}$, and therefore the thickness of the adjustment region around the front $Z=Z_C(\tau)$ scales like $(1-\ell)$. 

\subsubsection{Flow through the walls}
\begin{figure}
    \centering
    \includegraphics[width=0.85\linewidth]{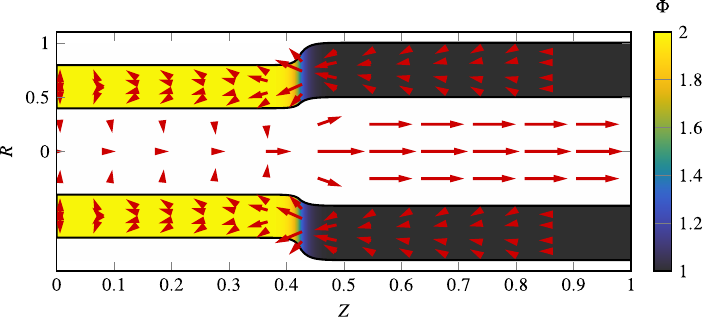}
    \caption{A plot at $\tau = 0.02$ of a drying gel tube with the same parameters as in figure \ref{fig:profiles}. The colours represent the polymer fraction field, with arrows in the gel showing the direction and magnitude of the interstitial flow field $\boldsymbol{u_g}$, as defined in equation \eqref{eqn:ug}. The arrows within the lumen show the flow within the tube, with the form of equation \eqref{eqn:lumen}.}
    \label{fig:flow_plot}
\end{figure}
Flow in the walls of the tube is driven by diffusive transport of water from more swollen regions to drier regions, with an interstitial fluid velocity
\begin{equation}
    \boldsymbol{u_g} = \frac{D(\phi)}{\phi}\bnabla\phi = \frac{k \Pi_{00}}{\mu_l}\left(\frac{1}{L}\pdv{\Phi_1}{Z}\boldsymbol{\hat{z}} + \frac{\varepsilon^2}{a_1}\pdv{\Phi_2}{R}\boldsymbol{\hat{r}}\right) \times
    \begin{dcases}
        \frac{\tilde{\Pi}}{\Phi_\infty} + \frac{4\mathcal{M}}{3}\Phi_1^{-2/3} \; &Z<Z_C \\
        1 + \frac{4\mathcal{M}}{3}\Phi_1^{-2/3} \; &Z>Z_C
    \end{dcases},
    \label{eqn:ug}
\end{equation}
at leading order in the aspect ratio. We define a dimensionless radial fluid velocity $U_g$ scaled with $a_1$ divided by the poroelastic timescale and an axial velocity $V_g$ scaled with $L$ divided by the same timescale, so that
\begin{equation}
    \left(V_g,\,U_g\right) = \left(\pdv{\Phi_1}{Z},\,\pdv{\Phi_2}{R}\right) \times
    \begin{dcases}
        \frac{\tilde{\Pi}}{\Phi_\infty} + \frac{4\mathcal{M}}{3}\Phi_1^{-2/3} \; &Z<Z_C. \\
        1 + \frac{4\mathcal{M}}{3}\Phi_1^{-2/3} \; &Z>Z_C.
    \end{dcases}
\end{equation}
Figure \ref{fig:flow_plot} illustrates an example flow field through the walls of the gel, with flow from more swollen to less swollen regions. In the dried region behind the temperature front, radial fluxes are outwards as water is driven out of the shrinking gel, with fluid transported axially towards the drier regions to the left. In $Z>Z_C$, however, fluxes are inwards towards the gel. In order to understand why this is, notice that the gel is more swollen on its interior than exterior when $Z<Z_C$ (as the tube fully dries from the outside in) and more swollen on its exterior than interior when $Z>Z_C$ (as the tube is fully swollen on $R=B_1$ with some loss of fluid in the interior due to axial fluxes towards the drier tube). Hence, there needs to be water drawn in from the surrounding fluid to replenish these regions.
\begin{figure}
    \centering
    \includegraphics[width=0.9\linewidth]{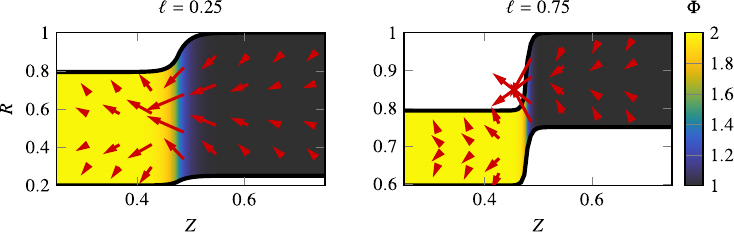}
    \caption{Plots close to $Z=Z_C(\tau)$ when $\tau = 0.025$, illustrating dominant radial flows when the gel is thinner ($\ell=0.75$) versus the thicker ($\ell = 0.25$) gel. In all other regards, the parameters are the same as in figure \ref{fig:flow_plot}. Notice the directional change either side of the drying front.}
    \label{fig:flow_plot_close}
\end{figure}

In general, therefore, the tube draws water inwards ahead of the deswelling front, and then expels the water behind this front. This is shown in detail in figure \ref{fig:flow_plot_close}, where the dominance of radial fluxes in thinner gel layers is also clear.

\subsubsection{Flow in the lumen}
There is also a flow of water through the lumen of the tube, resulting from mass conservation and the redistribution of fluid as the tube dries out. Assuming that the flow within the tube lumen can be described by a Stokes flow $\boldsymbol{v}=v_r\boldsymbol{\hat{r}} + v_z\boldsymbol{\hat{z}}$ with viscosity $\mu_l$,
\begin{equation}
    \mu_l\nabla^2 \boldsymbol{v} - \bnabla p = \boldsymbol{0} \qq{and} \bnabla \bcdot \boldsymbol{v} = 0.
\end{equation}
Using the slenderness approximation and assuming that pervadic pressure is independent of $r$, axial derivatives can be neglected in the radial component of the momentum equation, and it is found that the radial component of the velocity must be linear in $r$ if it is to be regular at $r=0$, and so
\begin{equation}
    v_r = C(z) r \qq{and} v_z = -2\int_0^z{C(z')\,\mathrm{d}z'},
\end{equation}
assuming that $v_z = 0$ at $z=0$ by symmetry. We then non-dimensionalise to find a radial velocity $U$ scaled with $a_1$ divided by the poroelastic timescale, and an axial velocity $V$ scaled with $L$ divided by the poroelastic timescale. The radial velocities in the gel are given by
\begin{equation}
    U_g = \frac{8\Phi_1^{2/3}\mathcal{D}(\Phi_1)\left[\Phi_0(T)-\Phi_1\right]}{\varepsilon^2(1-\ell)^2}\left(R - \frac{1+\ell}{2}\Phi_1^{-1/3}\right),
\end{equation}
and therefore, matching velocities at $R=\ell \Phi_1^{-1/3}$,
\begin{equation}
    U = -\frac{4\Phi_1\mathcal{D}\left[\Phi_0(T)-\Phi_1\right]}{\varepsilon^2(1-\ell)^2}R \qq{and} V = \frac{8}{\varepsilon^2(1-\ell)^2}\int_0^Z{\Phi_1\mathcal{D}\left[\Phi_0(T)-\Phi_1\right]\,\mathrm{d}Z'}.
    \label{eqn:lumen}
\end{equation}
Behind the temperature front, flow is radially inwards, since the gel dries into its interior, and this drives a forwards flow along the tube by mass conservation. The magnitude of this flow decreases ahead of the front as there is a weak outward radial flow here.

\begin{figure}
    \centering
    \subcaptionbox{Plots of the axial flow velocity at different times $\tau = 10^{-4}$ to $\tau = 10^{-1}$ when $A=100$\label{fig:pulses}}[0.49\textwidth]
    {\vspace{0.1cm}\includegraphics[width=0.94\linewidth]{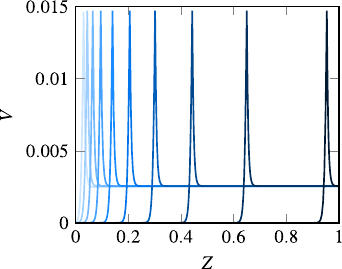}}
    \subcaptionbox{The maximum velocity $V_{\text{max}}$ when $A$ is varied, showing $V_{\text{max}} \sim A^{-1} \sim 1-\ell$\label{fig:vmax_vs_a}}[0.49\textwidth]{\vspace{0.08cm}\includegraphics[width=0.96\linewidth]{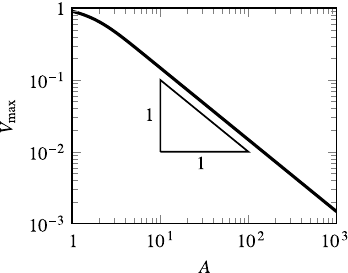}}
    \caption{Plots showing the approximate axial velocity $V$ (computed using the form of $\Phi_1$ in equation \eqref{eqn:tanh_fit}) for the parameters in table \ref{tab:params_tube}, showing how fluid travels in a pulse from the left to the right, with the height of the pulse inversely proportional to the fit parameter $A(\ell)$.}
    \label{fig:axial_flow}
\end{figure}

We can gain some insight into the nature of the axial fluid transport $V$ by considering the fitted form of equation \eqref{eqn:tanh_fit}, from which we can calculate the evolution of velocity in time. Figure \ref{fig:axial_flow} shows how fluid, stationary far behind the temperature front, is driven in the same direction as the thermal front, with a maximum axial velocity $V_{\text{max}}$ at $Z=Z_C(\tau)$. The velocity the decays to a constant value in the tube ahead of the front, which is nonzero by mass conservation. Figure \ref{fig:vmax_vs_a} shows that the height of this axial flow pulse scales like $A(\ell)^{-1} \sim 1-\ell$. Since the thickness of this adjustment region scales like $A(\ell)$, the height of the pulse increases with $\ell$, but its width decreases with $\ell$, such that the total flux carried by the pulse is constant as $\ell$ is varied.

\subsection{Summary of results}
In this section, we have used the conceptually simple model for a responsive tubular pump to illustrate a number of results attainable using the responsive LENS formalism that can then be applied to the design of more complicated responsive hydrogel devices. First, we illustrated how the relative impermeability of hydrogels allows for the gel dynamics problem to be decoupled from the fluid flow within cavities (in this example, the lumen of the gel tube), so that the induced pumping flows can be straightforwardly deduced as an output of our modelling, as justified in appendix \ref{app:coupling}. This allowed us to construct an explicit model for the deswelling of responsive gels as the temperature is changed, quantifying exactly how this deswelling is more rapid for tubes that are thinner $\sim (1-\ell)^2$ and showing how response times can be tuned by varying $\ell$, the thickness parameter.

Furthermore, the assumption of slenderness and the separation of timescales between slower poroelastic deformation and faster transport of heat allows for decoupling between the thermal problem and the poroelastic problem, and so modelling the behaviour of a thermo-responsive tube to a time- and space-varying temperature field is possible in this framework. We have seen how a tube relaxes to a new equilibrium state around a thermal front, and can quantify the spatial structure of the adjustment region, which is smoother and less well-defined for thicker tubes than thinner tubes that approach the equilibrium faster. Perhaps most instructively, the LENS model gives clear expressions for the interstitial flow velocities both along the axis of the tube and out of the walls, as well as the velocities within the hollow lumen, permitting predictions of the nature of the axial pumping fluxes to be made when a tube is heated from one point.

\section{Conclusion}
In this paper, we have extended the linear-elastic-nonlinear-swelling model outlined in \citet{Webber2023AGels} to incorporate a temperature-dependent osmotic pressure that can reproduce this behaviour when the temperature is brought above the LCST threshold. The approach is generic, and in some sense agnostic of the type of stimulus, and as such our model may be readily extended to, for instance, pH-responsive hydrogels. 

We showed that the approach of the linear-elastic-nonlinear-swelling theory is able to reproduce the transient swelling or deswelling behaviour of thermo-responsive gels both qualitatively and quantitatively. By choosing functional forms for the osmotic pressure and shear modulus that fit the parameters used in \citet{Butler2022TheHydrogels}, we are able to use LENS to reproduce predictions from a full nonlinear Flory-Huggins approach, provided that no spinodal decomposition occurs. Our model also provides criteria for such phase separation to occur when the diffusivity -- a function of macroscopic osmotic pressure and shear modulus -- is negative, and dried and swollen gels can coexist adjacent to one another. In order to regularise solutions of the polymer fraction evolution equation in these cases, it is likely necessary to incorporate some kind of surface energy to penalise the formation of new surfaces \citep{Hennessy2020PhaseBoundary}, leading to Korteweg stresses at internal interfaces. The question of how to describe such an approach in the context of a LENS model remains a topic for future research, since the formation of sharp polymer fraction gradients is not permitted in LENS.

Some of the key applications of thermo-responsive hydrogels are hampered by the slow response times of such gels to changes in the ambient temperature. In general, hydrogel swelling or drying is a slow process, mediated by viscously-dominated interstitial flows through a low-permeability scaffold, with some gels taking hours or days to reach an equilibrium state \citep{Bertrand2016DynamicsGel}. This is clearly undesirable in microfluidic devices or actuators, and having a tunable response time to changes in temperature may be desirable for certain applications \citep{Maslen2023AMicroactuators}. In order to investigate the response time of simple gel structures, we have considered the case of a hollow tube of gel that can act like a displacement pump.

In this geometry, even though the axial dimension may be large, deformation timescales are set by the diffusion of water through the thin walls, so morphological changes can occur much more rapidly than they would in a solid gel. This occurs because the shrinkage of the outside of the tube is no longer rate-limited by the need to deform and drive fluid through the interior of the gel, since water can flow relatively unimpeded down the lumen of the tube. The transport of water through the pipe-like structure that results can be used as a proxy measure of the speed of response, with water being transported large distances surprisingly quickly as a thermal signal propagates.

In order to model these tubes, we made a slenderness approximation that the polymer fraction varies axially at leading order, with only small radial corrections as water is expelled from the gel as the critical temperature threshold is exceeded. This facilitated a mathematical treatment similar to that used for transpiration through cylinders in \citet{Webber2023AFormulation}, and thus we can write down analytical expressions for all of the interstitial fluid fluxes in the gel and in the lumen. This approach permits us to tune the geometry of the tubes to match the exact response times desired, and allows for the computation of fluid flows through the pore matrix, along the axis of the tube, and out of the side walls.

Though there is no definitive measure of `response time' in more complex geometries, we have discussed how varying the geometry and material properties of the gel that forms the tube lining can affect the speed at which fluid is transported through the lumen and the sharpness of the fluid pulse at the deswelling front. As one might expect, it is seen that thinner tubes react more rapidly to changes in temperature, and also that the resultant fluid pulse is more spatially localised around the thermal pulse in such cases. We have also elucidated the dependence of the fluid pulse driven down the pump on both the osmotic and elastic properties of the material forming the tube, enabling the design of displacement pumps with specific response characteristics.

In the future, these simple model tubes could be connected together to form a network, propagating information about external stimuli through the medium of fluid pulses much more rapidly than in a solid block of hydrogel, forming the basis for a porous sponge built from porous hydrogel, with the pore size and geometry designed to match the desired material properties. This approach has already been taken experimentally in the design of microfluidic devices that exhibit dynamic anisotropy \citep{Maslen2023AMicroactuators}, and we hope that our modelling will provide potential qualitative insights into the design characteristics of such devices in the future. 

\backsection[Acknowledgements]{JJW thanks Matthew Hennessy and Matthew Butler for helpful discussions on thermoelasticity, and Grae Worster for comments on a draft of the manuscript. We are thankful to the three anonymous reviewers whose helpful comments have led to a much-improved exposition of this work and a more careful discussion of heat transfer and boundary conditions on the inside of the gel tube.}

\backsection[Funding]{This work was supported by the Leverhulme Trust Research Leadership Award `Shape-Transforming Active Microfluidics’ (RL-2019-014) to TDMJ.}

\backsection[Declaration of interests]{The authors report no conflict of interest.}

\backsection[Author ORCIDs]{J. J. Webber, https://orcid.org/0000-0002-0739-9574; T. D. Montenegro-Johnson, https://orcid.org/0000-0002-9370-7720}

\appendix

\section{LENS material parameters from an energy-based approach}
\label{app:energy_based}
In \citet{Butler2022TheHydrogels}, the standard energy density function for a thermo-responsive hydrogel \citep{Cai2011MechanicsHydrogels} is used, following Flory-Huggins mixture theory and a neo-Hookean elastic model for the polymer chains,
\begin{equation}
    \mathcal{W} = \frac{k_B T}{2 \Omega_p}\left[\operatorname{tr}{\left(\mathsfbi{F_d}\mathsfbi{F_d}^{\!\!\mathrm{T}}\right)} - 3 + 2\operatorname{ln}{\phi}\right] + \frac{k_B T}{\Omega_f}\left[\frac{1-\phi}{\phi}\operatorname{ln}{(1-\phi)} + \chi(\phi,\,T)(1-\phi)\right],
    \label{eqn:app:helmholtz_free_energy}
\end{equation}
where $\mathsfbi{F_d}$ is the deformation gradient tensor measured relative to a fully-dry polymer. We can rewrite $\mathsfbi{F_d}$ in terms of $\mathsfbi{F}$, the deformation gradient measured relative to a state where $\phi\equiv\phi_{00}$, since the transition between the two states can be described by an isotropic scaling transformation,
\begin{equation}
    \mathsfbi{F_d} = \left(\phi_{00}^{-1/3}\mathsfbi{I}\right)\mathsfbi{F} = \phi_{00}^{-1/3}\mathsfbi{F} \qq{so} \operatorname{tr}{\left(\mathsfbi{F_d}\mathsfbi{F_d}^{\!\!\mathrm{T}}\right)} = \phi_{00}^{-2/3}\mathsf{F}_{ab}\mathsf{F}_{ab},
    \label{eqn:app:fd_vs_f}
\end{equation}
using Einstein summation convention. Following the approach of \citet{Cai2012EquationsGels}, the Terzaghi effective stress tensor $\mathsfbi{\sigma}^{(e)}$ (i.e. $\mathsfbi{\sigma}+p\mathsfbi{I}$) has components given by
\begin{equation}
    \mathsf{\sigma}^{(e)}_{ij} = \phi \pdv{\mathcal{W}}{\mathsf{F}_{ik}}\mathsf{F}_{jk},
    \label{eqn:app:stress_from_energy}
\end{equation}
again using summation convention. This derivation is based on the classical approach by \citet{Coleman1963TheViscosity}, coupling a local entropy imbalance law with the expression for the rate of change of internal energy \citep{Brunner2024NumericalHydrogels}. Since $\operatorname{det}\mathsfbi{F} = \phi_{00}/\phi$, the expression for the derivative of a determinant with respect to a matrix \citep{Petersen2012The2012} implies that
\begin{equation}
    \pdv{\phi}{\mathsf{F}_{ik}} = -\phi \mathsf{F}_{ki}^{-1}.
\end{equation}
Hence,
\begin{align}
    \pdv{\mathcal{W}}{\mathsf{F}_{ik}} &= \frac{k_B T}{\Omega_f}\left\lbrace \frac{1}{\Omega \phi_{00}^{2/3}}\mathsf{F}_{ik} + \left[\frac{\operatorname{ln}{(1-\phi)}}{\phi} + 1 +\phi\chi(\phi,\,T)-\phi(1-\phi)\pdv{\chi}{\phi} - \frac{1}{\Omega}\right]\mathsf{F}_{ki}^{-1}\right\rbrace \; \text{and} \notag \\
    \mathsf{\sigma}^{(e)}_{ij} &= \frac{k_B T}{\Omega_f}\left\lbrace\left[\operatorname{ln}{(1-\phi)} + \phi +\phi^2\chi-\phi^2(1-\phi)\pdv{\chi}{\phi} - \frac{\phi}{\Omega}\right]\delta_{ij} + \frac{\phi}{\Omega\phi_{00}^{2/3}}\mathsf{F}_{ik}\mathsf{F}_{jk} \right\rbrace,\!
\end{align}
where $\Omega = \Omega_p/\Omega_f$ represents the volume of polymer molecules relative to solvent molecules. Separating the deformation gradient into an isotropic part due to swelling and shrinkage and a deviatoric part that can be related to deviatoric Cauchy strain $\mathsfbi{\epsilon}$ \citep{Webber2023AGels},
\begin{equation}
    \mathsf{F}_{ik}\mathsf{F}_{jk} = \left(\frac{\phi}{\phi_{00}}\right)^{-2/3}\delta_{ij} + \frac{2\phi_{00}}{\phi}\mathsf{\epsilon}_{ij},
\end{equation}
and so the two temperature-dependent material parameters are
\begin{subequations}
    \begin{gather}
        \Pi(\phi) = \frac{k_B T}{\Omega_f}\left[\Omega^{-1}\left(\phi-\phi^{1/3}\right) - \phi - \operatorname{ln}{(1-\phi)} - \phi^2\chi+\phi^2(1-\phi)\pdv{\chi}{\phi}\right] \quad \text{and} \label{eqn:osmotic_pressure}\\
        \mu_s(\phi) = \frac{k_B T \phi_{00}^{1/3}}{\Omega_p}. \label{eqn:shear_modulus}
    \end{gather}
\end{subequations}
Notice that the shear modulus is independent of polymer fraction, and increases with temperature and chain length (longer polymer chains have a larger $\Omega_p$). The temperature-dependence of the osmotic pressure is more complicated, with contributions from the $k_B T$ prefactor, $\chi$, and $\pdv*{\chi}{\phi}$.

\begin{figure}
    \centering
    \subcaptionbox{ANB parameters}[0.49\textwidth]
    {\vspace{0.1cm}\includegraphics[width=0.95\linewidth]{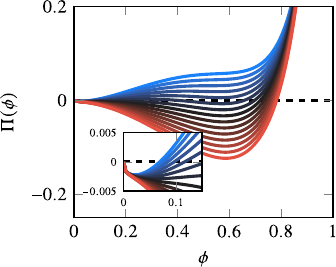}}
    \subcaptionbox{HHT parameters}[0.49\textwidth]{\vspace{0.1cm}\includegraphics[width=0.865\linewidth]{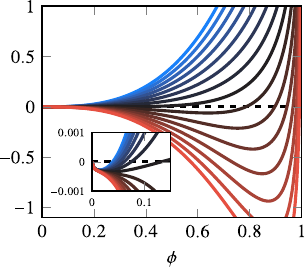}}
    \caption{Plots of the osmotic pressure \eqref{eqn:osmotic_pressure} for the two choices of fitted parameters in table \ref{tab:params} as the temperature is raised from $300\,\mathrm{K}$ (below $T_C$) (blue) to $315\,\mathrm{K}$ (above $T_C$) (red). Notice the change in equilibrium polymer fraction as the threshold is crossed.}
    \label{fig:osmotic_examples}
\end{figure}

To incorporate temperature dependence in $\chi(\phi,\,T)$, \citet{Butler2022TheHydrogels} specify an interaction parameter that depends linearly on both $\phi$ and $T$,
\begin{equation}
    \chi(\phi,\,T) = A_0 + B_0 T + (A_1+B_1 T)\phi,
\end{equation}
where the four parameters can be fitted to existing models in the literature. Here, we consider two example models -- the first is based on \citet{Afroze2000PhaseNetworks} (ANB), and the second is based on \citet{Hirotsu1987Volume-phaseGels} and henceforth referred to as HHT. The fitting parameters, as found in \citet{Butler2022TheHydrogels}, are summarised in table \ref{tab:params}. Figure \ref{fig:osmotic_examples} shows plots of the osmotic pressure in the case of these two parameter choices, showing (slightly) negative values of $\Pi$ as $\phi \to 0$ corresponding to states with a propensity to deswell, and $\Pi \to \infty$ as $\phi \to 1$, illustrating very dry states with a propensity to swell. As the temperature is increased, the location of equilibrium states changes.
\begin{table}
    \begin{center}
        \def~{\hphantom{0}}
        \begin{tabular}{lccccc}
            Model  & $A_0$  & $A_1$ & $B_0$ & $B_1$ & $\Omega$ \\[3pt]
            ANB \citep{Afroze2000PhaseNetworks} & $-12.947$ & $17.92$ & $0.04496\,\mathrm{K}^{-1}$ & $-0.0569\,\mathrm{K}^{-1}$ & $100$ \\
            HHT \citep{Hirotsu1987Volume-phaseGels} & $-62.22$ & $-58.28$ & $0.20470\,\mathrm{K}^{-1}$ & $0.19044\,\mathrm{K}^{-1}$ & $720$ \\
        \end{tabular}
        \caption{Fitted parameter values for the two thermo-responsive hydrogels considered in \citet{Butler2022TheHydrogels}, based on two pre-existing models from the literature.}
        \label{tab:params}
    \end{center}
\end{table}

\begin{figure}
    \begin{center}
        \includegraphics[width=0.95\linewidth]{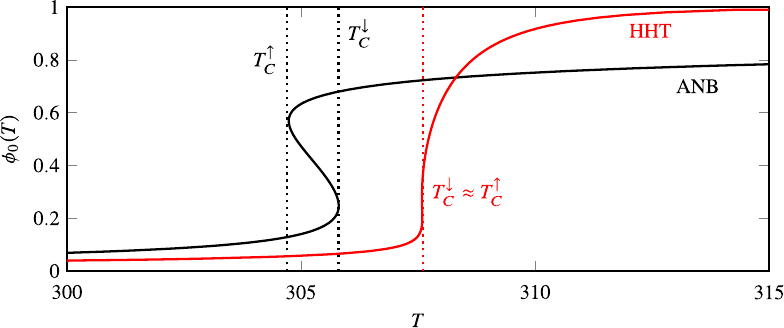}
        \caption{Plots of the equilibrium polymer fraction, determined by $\Pi(\phi_0) = 0$ in equation \eqref{eqn:phi_0_full}. Two choices of parameter values are plotted; those determined by \citet{Afroze2000PhaseNetworks} (ANB) and \citet{Hirotsu1987Volume-phaseGels} (HHT), showing the volume phase transition temperatures for swelling ($T_C^\uparrow$) and shrinking ($T_C^\downarrow$), respectively.}
        \label{fig:phi_0}
    \end{center}
\end{figure}
To find these equilibrium polymer fractions, we set $\Pi=0$ and thus consider the expression
\begin{equation}
    \Omega^{-1}\left(\phi_0-\phi_0^{1/3}\right) - \phi_0 - \operatorname{ln}{(1-\phi_0)} - \phi_0^2\left[A_0 + B_0 T + (2\phi_0-1)(A_1 + B_1 T)\right] = 0,
    \label{eqn:phi_0_full}
\end{equation}
for the two choices of parameters, and figure \ref{fig:phi_0} shows the variation of $\phi_0$ with temperature in both the ANB and HHT parameter sets. In the case of the parameters of \citet{Afroze2000PhaseNetworks}, it is especially apparent that there are two critical temperatures. As the temperature is lowered from around $310\,\mathrm{K}$, and the equilibrium polymer fraction $\phi_0$ decreases (swelling), there is a rapid increase in swelling at $T_C^\uparrow \approx 304.5\,\mathrm{K}$, the swelling critical temperature. As the temperature is increased from around $300\,\mathrm{K}$, however, there is a different critical temperature, $T_C^\downarrow \approx 306\,\mathrm{K}$ at which there is rapid drying.
This hysteresis is in fact exhibited in the case of both sets of parameters, where there are multiple solutions in a narrow band of temperatures around the critical volume phase transition temperature $T_C$, an effect which we ignore in the present study, modelling the equilibrium polymer fraction as single-valued at any temperature.

In the low-temperature (i.e. swollen) states, we further assume that $\phi_0 \ll 1$, so the leading-order balance of equation \eqref{eqn:phi_0_full} is
\begin{equation}
    \phi_0 \approx \left[\Omega\left(\frac{1}{2} - (A_0-A_1) - (B_0 - B_1)T\right)\right]^{-3/5}, 
\end{equation}
equal to the classical approximation in gels that are not thermo-responsive \citep{Doi2009GelDynamics, Webber2023AGels}. In both of the models, this gives $\phi_0 \sim 0.01$ for sufficiently low temperatures, but there is a singularity at
\begin{equation}
    T = \frac{1 - 2(A_0-A_1)}{2(B_0-B_1)},
\end{equation}
where the assumption of small polymer fraction can no longer be applied, corresponding to approximately $308\,\mathrm{K}$ in the ANB model and $311\,\mathrm{K}$ in the HHT model. This is close to the measured critical temperatures at which the affinity for water molecules drops rapidly and the gel dries out, $T_C$ (equal to around $305\,\mathrm{K}$ and $307.6\,\mathrm{K}$ in the two cases, respectively).

\section{Quantifying the coupling between lumen flow and gel dynamics}
\label{app:coupling}
Boundary conditions on the exterior of the tube are determined based on the assumption that $p\equiv 0$ in the quiescent fluid surrounding the hydrogel tube, a reasonable assumption in an unbounded fluid bath. However, assuming that $p\equiv 0$ on the interior of the tube, where we would instead expect pressure gradients to balance viscous stresses resulting from lumen fluxes, is not a valid approach to seeking an expression for the polymer fraction at $r=b_0(z,\,t)$. 

Making the assumption that pressures and stresses are independent of scaled radial position $R=r/a_1$ and depend only on the distance along the tube axis, we still impose $\mathsf{\sigma}_{rr} = 0$ at $R = B_0 = b_0/a_1$. We must solve for the Stokes flow inside the lumen that couples to the dynamics of the tube through a mixed boundary condition. Describing this flow by $\boldsymbol{v} = v_r\boldsymbol{\hat{r}} + v_z\boldsymbol{\hat{z}}$,
\begin{equation}
    \mu_l \nabla^2 \boldsymbol{v} - \bnabla p = \boldsymbol{0} \qq{and} \bnabla \bcdot \boldsymbol{v} = 0,
    \label{eqn:lumen_ns_eqn}
\end{equation}
where $\mu_l$ is the dynamic viscosity of the water in the tube. Since $p = p(Z)$, the radial component of this equation gives
\begin{equation}
    \frac{1}{R}\pdv{R}\left(R\pdv{v_r}{R}\right) - \frac{v_r}{R^2} + \varepsilon^2 \pdv[2]{v_r}{Z} = 0,
\end{equation}
which, at leading order in $\varepsilon$, is solved by $v_r = C(Z)R$ (requiring regularity at $R=0$). Thence, incompressibility allows us to find the form of $v_z$,
\begin{equation}
    \pdv{v_z}{Z} + \frac{1}{\varepsilon R}\pdv{R}\left(Rv_r\right) = 0 \qq{so} v_z = -\frac{2}{\varepsilon}\int_0^Z{C(z')\,\mathrm{d}z'}.
\end{equation}
This shows that, as expected, axial flows are much faster than radial flows, owing to slenderness. Substituting $v_z$ into the axial component of equation \eqref{eqn:lumen_ns_eqn} shows that
\begin{equation}
    \pdv{p}{Z} = -\frac{2\mu_l}{a_1}\pdv{C}{Z} \qq{so} p = -\frac{2\mu_l}{a_1}C(Z),
\end{equation}
on the assumption that $p=0$ in regions where there is no deswelling, i.e. where $C=0$ and there is no flow. Hence, since $p+\Pi(\phi) = 0$ at the inner tube surface,
\begin{equation}
    \phi_2 = \frac{\phi_0(T)-\phi_1}{\varepsilon^2} + \frac{2\mu_l}{\varepsilon^2 a_1 \Pi_0(T)}C(Z) \quad \text{at $R=B_0$.}
    \label{eqn:phi2_bc_basic}
\end{equation}
Notice that this forces both $C(Z)$ (the magnitude of the pervadic pressure corrections) and $\phi_0-\phi_1$ to be order $\varepsilon^2$, corresponding to a tube where the polymer fraction is everywhere close to its equilibrium value. To find $C(Z)$, we match radial fluid velocities in the gel and the lumen,
\begin{equation}
    v_r = \left.\frac{D(\phi,\,T)}{a_1 \phi}\pdv{\phi}{R}\right|_{R=B_0} \qq{so} C(Z) = \frac{\varepsilon^2 D(\phi_1,\,T)}{a_1 \phi_1 B_0}\left.\pdv{\phi_2}{R}\right|_{R=B_0},
\end{equation}
resulting in order-$\varepsilon^2$ corrections to the pervadic pressure field on the inside of the tube, as expected. This combines with equation \eqref{eqn:phi2_bc_basic} to give a mixed boundary condition on $\phi_2$ at $R=B_0$,
\begin{equation}
    \phi_2 - \frac{2 \mu_l D(\phi_1,\,T)}{a_1^2 \phi_1 \Pi_0(T) B_0}\left.\pdv{\phi_2}{R}\right|_{R=B_0} = \frac{\phi_0-\phi_1}{\varepsilon^2}.
    \label{eqn:app:full_interior_bc}
\end{equation}
To understand the relative importance of terms on the left-hand side of this boundary condition, introduce a non-dimensional coupling parameter $G$,
\begin{equation}
    G \sim \frac{\mu_l D}{a_1^2 \Pi_0} \sim \frac{k}{a_1^2},
\end{equation}
scaling the diffusivity using its form in equation \eqref{eqn:tube_evolution}, $D \sim k \Pi_{00}/\mu_l$. Since we expect $k \sim 10^{-15}\,\mathrm{m}^2$ \citep{Etzold2021TranspirationHydrogels}, it is reasonable to assume $G \ll 1$ for all tubes of relaxed radius $a_1 \gtrsim 10^{-7}\,\mathrm{m}$ -- hence it is possible to uncouple the interior flow from the dynamics of the inner surface of the tube, and we can assume that the same boundary condition $\phi_2 = [\phi_0(T)-\phi_1]/\varepsilon^2$ holds on the interior as on the exterior.

\section{Thermoelasticity and heat transfer in gels}\label{app:thermoelasticity}
All hyperelastic models based on an energy density function $\mathcal{W}$ (the Helmholtz free energy) require an approach based on thermodynamics to derive the components of the stress tensor in terms of the deformation \citep{Zaoui2001Elasticity:Treatment}. A number of recent works have sought models that couple chemical diffusion, thermodynamics and swelling to model thermo-responsive gels, but these are usually formulated in terms of energy density functions and not the Eulerian continuum-mechanical quantities in this study \citep{Brunner2024NumericalHydrogels}. Following a standard approach pioneered by \citet{Coleman1963TheViscosity}, and detailed in \citet{Chester2011ThermoGels} and \citet{Drozdov2014SwellingHydrogels}, we can write down an expression for the internal energy of a hydrogel per unit volume,
\begin{equation}
    \dv{U_d}{t} = R_d - \bnabla_{\boldsymbol{d}} \bcdot \boldsymbol{Q_d} + \mathsfbi{P}\mathsfbi{\colon}\dv{\mathsfbi{F_d}}{t} + \mu \dv{C_d}{t} - \boldsymbol{J_d}\bcdot\bnabla_{\boldsymbol{d}} \mu,
    \label{eqn:app:lagrangian_energy_balance}
\end{equation}
where $R_d$ is the external supply of heat per unit volume, $C_d$ is the number density of water molecules per unit reference volume, $\mu$ is the chemical potential, $\boldsymbol{Q_d}$ is the heat flux, $T$ is the temperature, $\boldsymbol{J_d}$ is the flux of water molecules, $\mathsfbi{F_d}$ is the deformation gradient tensor from a dry reference state and $\mathsfbi{P}$ is the first Piola-Kirchhoff stress tensor. The subscript ${}_d$ references quantities measured relative to a `dry' reference state where $\phi\equiv 1$ (as detailed, for example, in equation \eqref{eqn:app:fd_vs_f}).

In an Eulerian reference frame, it is standard to take Fourier's law of conduction to describe the heat flux $\boldsymbol{Q}$, and the molecular flux of water $\boldsymbol{J}$ is simply found by dividing the relative fluid volume flux by the volume of a single water molecule $\Omega_f$, such that
\begin{equation}
    \boldsymbol{Q} = - \rho c \kappa \bnabla T \qq{and} \boldsymbol{J} = -\frac{k(\phi)}{\mu_l \Omega_f}\bnabla p,
\end{equation}
where $\rho$ is the material density, $c$ is the specific heat capacity and $\kappa$ is the thermal diffusivity. Furthermore, an expression for $C_d$ can be found by appealing to incompressibility of water and polymer phases, such that any increases in volume from the dry state are due to the addition of water molecules alone, and hence
\begin{equation}
    C_d = \frac{\phi^{-1}-1}{\Omega_f} \qq{so} \dv{C_d}{t} = -\frac{1}{\Omega_f \phi^2}\dv{\phi}{t}.
\end{equation}

In order to rewrite the energy balance of equation \eqref{eqn:app:lagrangian_energy_balance} in an Eulerian form instead of its usual Lagrangian form with reference to a dry state, we make use of the assumption of LENS modelling that all deformation is, at leading order, isotropic. Hence,
\begin{equation}
    \mathsfbi{F} \approx \left(\frac{\phi}{\phi_0}\right)^{-1/3}\mathsfbi{I} \qq{so} \mathsfbi{F_d} \approx \phi^{-1/3}\mathsfbi{I},
\end{equation}
making use of the relation between $\mathsfbi{F}$ and $\mathsfbi{F_d}$ presented in equation \eqref{eqn:app:fd_vs_f}. This governing assumption also allows us to replace all instances of $\bnabla_{\boldsymbol{d}}$ with $\phi^{-1/3}\bnabla$ at leading order in the deviatoric strain, since the Eulerian state is approximately an isotropic dilatation of the fully-dry polymer reference state. The Piola-Kirchhoff and Cauchy strains are related via
\begin{equation}
    \mathsfbi{P} = \phi^{-1}\mathsfbi{\sigma}\mathsfbi{F_d}^{-\mathrm{T}} \approx \phi^{-2/3}\mathsfbi{\sigma} \qq{hence} \mathsfbi{P}\mathsfbi{\colon}\dv{\mathsfbi{F_d}}{t} \approx -\frac{1}{3\phi^2}\dv{\phi}{t}\operatorname{tr}{\mathsfbi{\sigma}}.
\end{equation}
Finally, since lengths scale with $\phi^{-1/3}$ from the dry state to the swollen state, and volumes correspondingly by $\phi^{-1}$,
\begin{equation}
    U = \phi U_d,\quad R = \phi R_d,\quad\boldsymbol{Q} = \phi^{2/3}\boldsymbol{Q_d} \qq{and} \boldsymbol{J} = \phi^{2/3}\boldsymbol{J_d},
\end{equation}
and hence equation \eqref{eqn:app:lagrangian_energy_balance} can be rewritten as
\begin{align}
    \dv{t}\left(\frac{U}{\phi}\right) &= \frac{R}{\phi} - \phi^{-1/3}\bnabla\bcdot\left(\phi^{-2/3}\boldsymbol{Q}\right)-\frac{\mu}{\Omega_f \phi^2}\dv{\phi}{t} - \frac{1}{\phi}\boldsymbol{J}\bcdot\bnabla\mu -\frac{1}{3\phi^2}\dv{\phi}{t}\operatorname{tr}{\mathsfbi{\sigma}} \notag \\
    &= \frac{R}{\phi} + c \kappa \phi^{-1/3}\bnabla\bcdot\left(\phi^{-2/3}\bnabla T\right)-\frac{\mu}{\Omega_f \phi^2}\dv{\phi}{t} + \frac{k(\phi)}{\mu_l \Omega_f \phi}\bnabla p \bcdot \bnabla \mu -\frac{1}{3\phi^2}\dv{\phi}{t}\operatorname{tr}{\mathsfbi{\sigma}}.
\end{align}
Expanding all terms of this equation and noting that the pervadic pressure and chemical potential are related by $p=\mu/\Omega_f$ alongside $\operatorname{tr}{\mathsfbi{\sigma}} = -3(p+\Pi)$,
\begin{equation}
    \dv{t}\left(\frac{U}{\phi}\right) = \frac{R}{\phi} + \frac{\rho c \kappa}{\phi}\nabla^2 T - \frac{2 \rho c \kappa}{3\phi^2}\bnabla\phi \bcdot \bnabla T + \frac{k(\phi)}{\mu_l \phi}\left|\bnabla p\right|^2 + \frac{\Pi(\phi)}{\phi^2}\dv{\phi}{t}.
\end{equation}
Now, if the internal energy is given by $\rho c T$ and the added assumption that density remains approximately constant in time is made,
\begin{equation}
    \dv{T}{t} = \frac{R}{\rho c} + \kappa \nabla^2 T - \frac{2 \kappa}{3 \phi}\bnabla \phi \bcdot \bnabla T + \frac{k(\phi)}{\rho c \mu_l}\left|\bnabla p\right|^2 + \frac{1}{\phi}\left(\frac{\Pi(\phi)}{\rho c} + T\right)\dv{\phi}{t}.
\end{equation}
LENS scalings show that gradients in polymer fraction are small on the order of the deviatoric strain, and therefore the $\bnabla \phi \bcdot \bnabla T$ term is much smaller than that featuring $\nabla^2 T$. Furthermore, we can replace the total derivatives with the material derivative advecting with the deformation of the gel itself, so the leading order temperature evolution equation is
\begin{equation}
    \pdv{T}{t} + \boldsymbol{q}\bcdot\bnabla T = \frac{R}{\rho c} + \kappa \nabla^2 T + \frac{k(\phi)}{\rho c \mu_l}\left|\bnabla p\right|^2 + \frac{1}{\phi}\left(\frac{\Pi(\phi)}{\rho c} + T\right)\left(\pdv{\phi}{t} + \boldsymbol{q}\bcdot\bnabla\phi\right).
\end{equation}
This heat equation shows how temperature evolves due to material fluxes (the advective term), diffusion, and then two additional terms related to swelling and internal flows. The first term is equal to $(\boldsymbol{u}\bcdot\boldsymbol{F})/\rho c$ where $\boldsymbol{F}=-\bnabla p$ and so represents the rate of work done by the pervadic pressure gradients in the interstitial flow. The second, related to the osmotic pressure and changes in polymer fraction, arises as energy is either used up or released as water molecules associate and dissociate with polymer chains, and can hence be seen as an analogue of latent heat in phase change.

\bibliographystyle{jfm}
\bibliography{bib}

\begin{thebibliography}{44}
\expandafter\ifx\csname natexlab\endcsname\relax\def\natexlab#1{#1}\fi
\def\au#1{#1} \def\ed#1{#1} \def\yr#1{#1}\def\at#1{#1}\def\jt#1{\textit{#1}}
  \def\bt#1{#1}\def\bvol#1{\textbf{#1}} \def\vol#1{#1} \def\pg#1{#1}
  \def\publ#1{#1}\def\arxiv#1{#1}\def\org#1{#1}\def\st#1{\textit{#1}}

\bibitem[Abramowitz \& Stegun(1970)]{Abramowitz1970HandbookSeries}
{\sc \au{Abramowitz, M.} \& \au{Stegun, I.~A.}} \yr{1970} {\em {Handbook of
  Mathematical Functions: with Formulas, Graphs, and Mathematical Tables}\/}.
  \publ{National Bureau of Standards}.

\bibitem[Afroze {\em et~al.\/}(2000)Afroze, Nies \&
  Berghmans]{Afroze2000PhaseNetworks}
{\sc \au{Afroze, F.}, \au{Nies, E.} \& \au{Berghmans, H.}} \yr{2000}
  \at{{Phase transitions in the system poly(N-isopropylacrylamide)/water and
  swelling behaviour of the corresponding networks}}.  \jt{J. Mol. Struct.}
  \bvol{554}~(1),  \pg{55--68}.

\bibitem[Bertrand {\em et~al.\/}(2016)Bertrand, Peixinho, Mukhopadhyay \&
  MacMinn]{Bertrand2016DynamicsGel}
{\sc \au{Bertrand, T.}, \au{Peixinho, J.}, \au{Mukhopadhyay, S.} \&
  \au{MacMinn, C.~W.}} \yr{2016}  \at{{Dynamics of swelling and drying in a
  spherical gel}}.  \jt{Phys. Rev. Appl.}  \bvol{6}~(6),  \pg{064010}.

\bibitem[Brunner {\em et~al.\/}(2024)Brunner, Seidlhofer \&
  Ulz]{Brunner2024NumericalHydrogels}
{\sc \au{Brunner, F.}, \au{Seidlhofer, T.} \& \au{Ulz, M.~H.}} \yr{2024}
  \at{{A numerical model for chemo-thermo-mechanical coupling at large strains
  with an application to thermoresponsive hydrogels}}.  \jt{Comput. Mech.}
  \bvol{74},  \pg{509--536}.

\bibitem[Butler \& Montenegro-Johnson(2022)]{Butler2022TheHydrogels}
{\sc \au{Butler, M.~D.} \& \au{Montenegro-Johnson, T.~D.}} \yr{2022}  \at{{The
  swelling and shrinking of spherical thermo-responsive hydrogels}}.  \jt{J.
  Fluid Mech.}  \bvol{947},  \pg{A11}.

\bibitem[Cai \& Suo(2011)]{Cai2011MechanicsHydrogels}
{\sc \au{Cai, S.} \& \au{Suo, Z.}} \yr{2011}  \at{{Mechanics and chemical
  thermodynamics of phase transition in temperature-sensitive hydrogels}}.
  \jt{J. Mech. Phys. Solids}  \bvol{59}~(11),  \pg{2259--2278}.

\bibitem[Cai \& Suo(2012)]{Cai2012EquationsGels}
{\sc \au{Cai, S.} \& \au{Suo, Z.}} \yr{2012}  \at{{Equations of state for ideal
  elastomeric gels}}.  \jt{EPL}  \bvol{97}~(3),  \pg{34009}.

\bibitem[Chester \& Anand(2011)]{Chester2011ThermoGels}
{\sc \au{Chester, S.~A.} \& \au{Anand, L.}} \yr{2011}  \at{{A
  thermo-mechanically coupled theory for fluid permeation in elastomeric
  materials: Application to thermally responsive gels}}.  \jt{J. Mech. Phys.
  Solids}  \bvol{59},  \pg{1978--2006}.

\bibitem[Coleman \& Noll(1963)]{Coleman1963TheViscosity}
{\sc \au{Coleman, B.~D.} \& \au{Noll, W.}} \yr{1963}  \at{{The thermodynamics
  of elastic materials with heat conduction and viscosity}}.  \jt{Arch. Ration.
  Mech. Anal.}  \bvol{13}~(1),  \pg{167--178}.

\bibitem[Curatolo {\em et~al.\/}(2023)Curatolo, Lisi, Napoli \&
  Nardinocchi]{Curatolo2023CircumferentialDehydration}
{\sc \au{Curatolo, M.}, \au{Lisi, F.}, \au{Napoli, G.} \& \au{Nardinocchi, P.}}
  \yr{2023}  \at{{Circumferential buckling of a hydrogel tube emptying upon
  dehydration}}.  \jt{Eur. Phys. J. Plus}  \bvol{138},  \pg{382}.

\bibitem[Curatolo {\em et~al.\/}(2018)Curatolo, Nardinocchi \&
  Teresi]{Curatolo2018DrivingCavity}
{\sc \au{Curatolo, M.}, \au{Nardinocchi, P.} \& \au{Teresi, L.}} \yr{2018}
  \at{{Driving water cavitation in a hydrogel cavity}}.  \jt{Soft Matt.}
  \bvol{14},  \pg{2310--2321}.

\bibitem[Das {\em et~al.\/}(2015)Das, Ghosh, Ghosh, Haldar, Dhara, Panda \&
  Pal]{Das2015StimulusRelease}
{\sc \au{Das, D.}, \au{Ghosh, P.}, \au{Ghosh, A.}, \au{Haldar, C.}, \au{Dhara,
  S.}, \au{Panda, A.~B.} \& \au{Pal, S.}} \yr{2015}  \at{{Stimulus-responsive,
  biodegradable, biocompatible, covalently cross-linked hydrogel based on
  dextrin and poly(N-isopropylacrylamide) for in vitro/in vivo controlled drug
  release}}.  \jt{ACS Appl. Mater. Interfaces}  \bvol{7},  \pg{14338--14351}.

\bibitem[Doi(2009)]{Doi2009GelDynamics}
{\sc \au{Doi, M.}} \yr{2009}  \at{{Gel dynamics}}.  \jt{J. Phys. Soc. Jpn.}
  \bvol{78}~(5),  \pg{052001}.

\bibitem[Dong \& Jiang(2007)]{Dong2007AutonomousHydrogels}
{\sc \au{Dong, L.} \& \au{Jiang, H.}} \yr{2007}  \at{{Autonomous microfluidics
  with stimuli-responsive hydrogels}}.  \jt{Soft Matter}  \bvol{3},
  \pg{1223--1230}.

\bibitem[Drozdov(2014)]{Drozdov2014SwellingHydrogels}
{\sc \au{Drozdov, A.~D.}} \yr{2014}  \at{{Swelling of thermo-responsive
  hydrogels}}.  \jt{EPJE}  \bvol{37}~(10),  \pg{93}.

\bibitem[Etzold {\em et~al.\/}(2021)Etzold, Linden \&
  Worster]{Etzold2021TranspirationHydrogels}
{\sc \au{Etzold, M.~A.}, \au{Linden, P.~F.} \& \au{Worster, M.~G.}} \yr{2021}
  \at{{Transpiration through hydrogels}}.  \jt{J. Fluid Mech.}  \bvol{925},
  \pg{A8}.

\bibitem[Freeman {\em et~al.\/}(1987)Freeman, Morgan \&
  Cullen]{Freeman1987ThermalChain}
{\sc \au{Freeman, J.~J.}, \au{Morgan, G.~J.} \& \au{Cullen, C.~A.}} \yr{1987}
  \at{{Thermal conductivity of a single polymer chain}}.  \jt{Phys. Rev. B}
  \bvol{35},  \pg{7627--7635}.

\bibitem[Gomez {\em et~al.\/}(2017)Gomez, Moulton \&
  Vella]{Gomez2017CriticalInstabilities}
{\sc \au{Gomez, M.}, \au{Moulton, D.} \& \au{Vella, D.}} \yr{2017}
  \at{{Critical slowing down in purely elastic `snap-through' instabilities}}.
  \jt{Nature Phys.}  \bvol{13},  \pg{142--145}.

\bibitem[Guilherme {\em et~al.\/}(2015)Guilherme, Aouada, Fajardo, Martins,
  Paulino, Davi, Rubira \& Muniz]{Guilherme2015SuperabsorbentReview}
{\sc \au{Guilherme, M.~R.}, \au{Aouada, F.~A.}, \au{Fajardo, A.~R.},
  \au{Martins, A.~F.}, \au{Paulino, A.~T.}, \au{Davi, M. F.~T.}, \au{Rubira,
  A.~F.} \& \au{Muniz, E.~C.}} \yr{2015}  \at{{Superabsorbent hydrogels based
  on polysaccharides for application in agriculture as soil conditioner and
  nutrient carrier: A review}}.  \jt{Eur. Polym. J.}  \bvol{72},
  \pg{365--385}.

\bibitem[Harmon {\em et~al.\/}(2003)Harmon, Tang \&
  Frank]{Harmon2003AHydrogels}
{\sc \au{Harmon, M.~E.}, \au{Tang, M.} \& \au{Frank, C.~W.}} \yr{2003}  \at{{A
  microfluidic actuator based on thermoresponsive hydrogels}}.  \jt{Polymer}
  \bvol{44}~(16),  \pg{4547--4556}.

\bibitem[Hennessy {\em et~al.\/}(2020)Hennessy, M{\"{u}}nch \&
  Wagner]{Hennessy2020PhaseBoundary}
{\sc \au{Hennessy, M.~G.}, \au{M{\"{u}}nch, A.} \& \au{Wagner, B.}} \yr{2020}
  \at{{Phase separation in swelling and deswelling hydrogels with a free
  boundary}}.  \jt{Phys. Rev. E}  \bvol{101}~(3),  \pg{032501}.

\bibitem[Hirotsu {\em et~al.\/}(1987)Hirotsu, Hirokawa \&
  Tanaka]{Hirotsu1987Volume-phaseGels}
{\sc \au{Hirotsu, S.}, \au{Hirokawa, Y.} \& \au{Tanaka, T.}} \yr{1987}
  \at{{Volume-phase transitions of ionized N-isopropylacrylamide gels}}.
  \jt{J. Chem. Phys.}  \bvol{87}~(2),  \pg{1392--1395}.

\bibitem[Kaviany(1995)]{Kaviany1995PrinciplesMedia}
{\sc \au{Kaviany, M.}} \yr{1995} {\em {Principles of Heat Transfer in Porous
  Media}\/}.  \publ{Springer New York}.

\bibitem[Lee {\em et~al.\/}(2020)Lee, Song \& Sun]{Lee2020HydrogelRobotics}
{\sc \au{Lee, Y.}, \au{Song, W.~J.} \& \au{Sun, J.~Y.}} \yr{2020}
  \at{{Hydrogel soft robotics}}.  \jt{Mater. Today Phys.}  \bvol{15},
  \pg{100258}.

\bibitem[Maslen {\em et~al.\/}(2023)Maslen, Gholamipour-Shirazi, Butler,
  Kropacek, Rehor \& Montenegro-Johnson]{Maslen2023AMicroactuators}
{\sc \au{Maslen, C.}, \au{Gholamipour-Shirazi, A.}, \au{Butler, M.~D.},
  \au{Kropacek, J.}, \au{Rehor, I.} \& \au{Montenegro-Johnson, T.~D.}}
  \yr{2023}  \at{{A new class of single-material, non-reciprocal
  microactuators}}.  \jt{Macromol. Rapid Commun.}  \bvol{44}~(6),
  \pg{2200842}.

\bibitem[Matsuo \& Tanaka(1988)]{Matsuo1988KineticsGels}
{\sc \au{Matsuo, E.~S.} \& \au{Tanaka, T.}} \yr{1988}  \at{{Kinetics of
  discontinuous volume–phase transition of gels}}.  \jt{J. Chem. Phys.}
  \bvol{89},  \pg{1695--1703}.

\bibitem[Neumann {\em et~al.\/}(2023)Neumann, di~Marco, Iudin, Viola, van
  Nostrum, van Ravensteijn \& Vermonden]{Neumann2023Stimuli-ResponsiveTomorrow}
{\sc \au{Neumann, M.}, \au{di~Marco, G.}, \au{Iudin, D.}, \au{Viola, M.},
  \au{van Nostrum, C.~F.}, \au{van Ravensteijn, B. G.~P.} \& \au{Vermonden,
  T.}} \yr{2023}  \at{{Stimuli-responsive hydrogels: the dynamic smart
  biomaterials of tomorrow}}.  \jt{Macromolecules}  \bvol{56}~(21),
  \pg{8377--8392}.

\bibitem[Nistane {\em et~al.\/}(2022)Nistane, Chen, Lee, Lively \&
  Ramprasad]{Nistane2022EstimationLearning}
{\sc \au{Nistane, J.}, \au{Chen, L.}, \au{Lee, Y.}, \au{Lively, R.} \&
  \au{Ramprasad, R.}} \yr{2022}  \at{{Estimation of the Flory-Huggins
  interaction parameter of polymer-solvent mixtures using machine learning}}.
  \jt{MRS Commun.}  \bvol{12},  \pg{1096--1102}.

\bibitem[Peppin {\em et~al.\/}(2005)Peppin, Elliott \&
  Worster]{Peppin2005PressureSuspensions}
{\sc \au{Peppin, S. S.~L.}, \au{Elliott, J. A.~W.} \& \au{Worster, M.~G.}}
  \yr{2005}  \at{{Pressure and relative motion in colloidal suspensions}}.
  \jt{Phys. Fluids}  \bvol{17}~(5),  \pg{053301}.

\bibitem[Petersen \& Pedersen(2012)]{Petersen2012The2012}
{\sc \au{Petersen, K.~B.} \& \au{Pedersen, M.~S.}} \yr{2012}  \at{{The Matrix
  Cookbook, November 2012}}.  \jt{Technical University of Denmark}
  \bvol{7}~(15).

\bibitem[Richter {\em et~al.\/}(2009)Richter, Klatt, Paschew \&
  Klenke]{Richter2009MicropumpsHydrogels}
{\sc \au{Richter, A.}, \au{Klatt, S.}, \au{Paschew, G.} \& \au{Klenke, C.}}
  \yr{2009}  \at{{Micropumps operated by swelling and shrinking of
  temperature-sensitive hydrogels}}.  \jt{Lab Chip}  \bvol{9},  \pg{613--618}.

\bibitem[Seo {\em et~al.\/}(2019)Seo, Wang, Chang, Park \&
  Kim]{Seo2019HydrogelRate}
{\sc \au{Seo, J.}, \au{Wang, C.}, \au{Chang, S.}, \au{Park, J.} \& \au{Kim,
  W.}} \yr{2019}  \at{{A hydrogel-driven microfluidic suction pump with a high
  flow rate}}.  \jt{Lab Chip}  \bvol{19},  \pg{1790--1796}.

\bibitem[Spratte {\em et~al.\/}(2022)Spratte, Arndt, Wacker, Hauck, Adelung,
  Schr\"oder, Sch\"utt \&
  Selhuber-Unkel]{Spratte2022ThermoresponsiveMicrochannels}
{\sc \au{Spratte, T.}, \au{Arndt, C.}, \au{Wacker, I.}, \au{Hauck, M.},
  \au{Adelung, R.}, \au{Schr\"oder, R.~R.}, \au{Sch\"utt, F.} \&
  \au{Selhuber-Unkel, C.}} \yr{2022}  \at{{Thermoresponsive hydrogels with
  improved actuation function by interconnected microchannels}}.  \jt{Adv.
  Intell. Sys.}  \bvol{4},  \pg{2100081}.

\bibitem[Tomari \& Doi(1995)]{Tomari1995HysteresisGels}
{\sc \au{Tomari, T.} \& \au{Doi, M.}} \yr{1995}  \at{{Hysteresis and incubation
  in the dynamics of volume transition of spherical gels}}.
  \jt{Macromolecules}  \bvol{28},  \pg{8334--8343}.

\bibitem[Vernerey \& Shen(2017)]{Vernerey2017TheEnvironment}
{\sc \au{Vernerey, F.} \& \au{Shen, T.}} \yr{2017}  \at{{The mechanics of
  hydrogel crawlers in confined environment}}.  \jt{J. R. Soc. Interface}
  \bvol{14}~(132),  \pg{20170242}.

\bibitem[Voudouris {\em et~al.\/}(2013)Voudouris, Florea, van~der Schoot \&
  Wyss]{Voudouris2013MicromechanicsLCST}
{\sc \au{Voudouris, P.}, \au{Florea, D.}, \au{van~der Schoot, P.} \& \au{Wyss,
  H.~M.}} \yr{2013}  \at{{Micromechanics of temperature sensitive microgels:
  dip in the Poisson ratio near the LCST}}.  \jt{Soft Matter}  \bvol{9},
  \pg{7158--7166}.

\bibitem[Webber(2024)]{Webber2024DynamicsHydrogels}
{\sc \au{Webber, J.~J.}} \yr{2024}  \at{{Dynamics of super-absorbent
  hydrogels}}. PhD thesis, University of Cambridge.

\bibitem[Webber {\em et~al.\/}(2023)Webber, Etzold \&
  Worster]{Webber2023AFormulation}
{\sc \au{Webber, J.~J.}, \au{Etzold, M.~A.} \& \au{Worster, M.~G.}} \yr{2023}
  \at{{A linear-elastic-nonlinear-swelling theory for hydrogels. Part 2.
  Displacement formulation}}.  \jt{J. Fluid Mech.}  \bvol{960},  \pg{A38}.

\bibitem[Webber \& Worster(2023)]{Webber2023AGels}
{\sc \au{Webber, J.~J.} \& \au{Worster, M.~G.}} \yr{2023}  \at{{A
  linear-elastic-nonlinear-swelling theory for hydrogels. Part 1. Modelling of
  super-absorbent gels}}.  \jt{J. Fluid Mech.}  \bvol{960},  \pg{A37}.

\bibitem[Xu {\em et~al.\/}(2018)Xu, Cai \& Liu]{Xu2018ThermalNanoscale}
{\sc \au{Xu, S.}, \au{Cai, S.} \& \au{Liu, Z.}} \yr{2018}  \at{{Thermal
  conductivity of polyacrylamide hydrogels at the nanoscale}}.  \jt{ACS Appl.
  Mater. Interfaces}  \bvol{10},  \pg{36352--36360}.

\bibitem[Xu {\em et~al.\/}(2024)Xu, Yue \& Feng]{Xu2024TheoryFlow}
{\sc \au{Xu, Z.}, \au{Yue, P.} \& \au{Feng, J.~J.}} \yr{2024}  \at{{A theory of
  hydrogel mechanics that couples swelling and external flow}}.  \jt{Soft
  Matt.}  \bvol{20},  \pg{5389--5406}.

\bibitem[Xu {\em et~al.\/}(2022)Xu, Zhang, Young, Yue \&
  Feng]{Xu2022ComparisonInterface}
{\sc \au{Xu, Z.}, \au{Zhang, J.}, \au{Young, Y.-N.}, \au{Yue, P.} \& \au{Feng,
  J.~J.}} \yr{2022}  \at{{Comparison of four boundary conditions for the
  fluid--hydrogel interface}}.  \jt{Phys. Rev. Fluids}  \bvol{7},  \pg{093301}.

\bibitem[Zaoui \& Stolz(2001)]{Zaoui2001Elasticity:Treatment}
{\sc \au{Zaoui, A.} \& \au{Stolz, C.}} \yr{2001}  \at{{Elasticity:
  Thermodynamic Treatment}}.  \jt{Encyclopedia of Materials: Science and
  Technology}  \pg{pp. 2445--2448}.

\bibitem[Zohuriaan-Mehr {\em et~al.\/}(2010)Zohuriaan-Mehr, Omidian, Doroudiani
  \& Kabiri]{Zohuriaan-Mehr2010AdvancesMaterials}
{\sc \au{Zohuriaan-Mehr, M.~J.}, \au{Omidian, H.}, \au{Doroudiani, S.} \&
  \au{Kabiri, K.}} \yr{2010}  \at{{Advances in non-hygienic applications of
  superabsorbent hydrogel materials}}.  \jt{J. Mater. Sci.}  \bvol{45}~(21),
  \pg{5711--5735}.

\end{thebibliography}

\end{document}